\documentclass[graybox]{svmult}

\usepackage{type1cm}        
\usepackage{makeidx}         
\usepackage{graphicx}        
\usepackage{multicol}        
\usepackage[bottom]{footmisc}
\usepackage{epstopdf} 

\usepackage{newtxtext}       %
\usepackage{newtxmath}       
\usepackage{bm}
\usepackage{natbib}
\makeindex             

\newcommand\bpsi{{\boldsymbol \psi}}
\newcommand\btheta{{\boldsymbol \theta}}

\newcommand\bGamma{{\boldsymbol \Gamma}}

\newcommand\bOmega{{\boldsymbol \Omega}}
\newcommand\bSigma{{\boldsymbol \Sigma}}
\newcommand\bsigma{{\boldsymbol \sigma}}
\newcommand\bXi{{\boldsymbol \Xi}}

\newcommand\bTheta{{\boldsymbol \Theta}}
\newcommand\bUpsilon{{\boldsymbol \Upsilon}}

\newcommand\x{\mathbf{x}} 
\newcommand\bI{\mathbf{I}}
\newcommand\bQ{\mathbf{Q}}

\newcommand\bR{\mathbf{R}}

\newcommand\bH{\mathbf{H}}
\newcommand\bL{\mathbf{L}}
\newcommand\bX{\mathbf{X}}

\newcommand\bC{\mathbf{C}}

\newcommand\bP{\mathbf{P}}
\newcommand\bK{\mathbf{K}}

\newcommand\bM{\mathbf{M}}

\newcommand\bU{\mathbf{U}}
\newcommand\bV{\mathbf{V}}
\newcommand\U{\mathcal{U}}

\newcommand\by{\mathbf{y}}
\newcommand\bw{\mathbf{w}}
\newcommand\bx{\mathbf{x}}
\newcommand\bv{\mathbf{v}}
\newcommand\bu{\mathbf{u}}

\newcommand\xa{\mathbf{x}^\mathrm{a}}

\newcommand\xf{\mathbf{x}^\mathrm{f}}

\newcommand{\T}{^\top}

\newcommand\bzero{{\mathbf 0}}

\renewcommand\vec{\mathop{\mathrm{vec}}\nolimits}

\newcommand\rank{\mathop{\mathrm{rank}}\nolimits}

\DeclareMathOperator*{\im}{\mathrm{Im}}

\renewcommand\({\left(}
\renewcommand\){\right)}
\newcommand\be{\begin{equation}}
\newcommand\ee{\end{equation}}

\newcommand\Nx{{N_\mathrm{x}}}
\newcommand\Ny{{N_\mathrm{y}}}

\newcommand\bone{\mathbf{1}}

\begin{document}
\title*{Data assimilation for chaotic dynamics}

\titlerunning{Short Title} 
\author{Alberto Carrassi, Marc Bocquet, Jonathan Demaeyer, Colin Grudzien, Patrick Raanes and Stephane Vannitsem }
\institute{Alberto Carrassi \at Dept. of Meteorology and National Centre for Earth Observation, University of Reading, Reading, UK,  \at Mathematical Institute, University of Utrecht, Netherlands
\email{n.a.carrassi@reading.ac.uk}
\and Marc Bocquet \at CEREA, joint laboratory \'Ecole des Ponts ParisTech
and EDF R\&D, Universit\`e Paris-Est, France \email{marc.bocquet@enpc.fr} 
\and Jonathan Demaeyer \at Royal Meteorological Institute of Belgium, Brussels, Belgium \email{jodemaey@meteo.be}
\and Colin Grudzien \at Department of Mathematics and Statistics, University of Nevada, Reno, Reno, Nevada, USA
\email{cgrudzien@unr.edu}
\and Patrick Raanes \at NORCE, Bergen, Norway
\email{para@norceresearch.no}
\and Stephane Vannitsem \at Royal Meteorological Institute of Belgium, Brussels, Belgium
\email{svn@meteo.be}}
%
%
\maketitle

\abstract{Chaos is ubiquitous in physical systems. The associated sensitivity to initial conditions is a significant obstacle in forecasting the weather and other geophysical fluid flows. Data assimilation is the process whereby the uncertainty in initial conditions is reduced by the astute combination of model predictions and real-time data. This chapter reviews recent findings from investigations on the impact of chaos on data assimilation methods: for the Kalman filter and smoother in linear systems, analytic results are derived; for their ensemble-based versions and nonlinear dynamics, numerical results provide insights. The focus is on 
characterizing the asymptotic statistics of the Bayesian posterior in terms of the dynamical instabilities, differentiating between deterministic and stochastic dynamics. We also present two novel results. Firstly, we study the functioning of the ensemble Kalman filter (EnKF) in the context of a chaotic, coupled, atmosphere-ocean model with a quasi-degenerate spectrum of Lyapunov exponents, showing the importance of having sufficient ensemble members to track all of the near-null modes. Secondly, for the fully non-Gaussian method of the particle filter, numerical experiments are conducted to test whether the curse of dimensionality can be mitigated by discarding observations in the directions of little dynamical growth of uncertainty.
The results refute this option, most likely because the particles already embody this information on the chaotic system.
The results also suggest that
it is the rank of the unstable-neutral subspace of the dynamics, and not that of the observation operator, that determines the required number of particles. We finally discuss how knowledge of the random attractor can play a role in the development of future data assimilation schemes for chaotic multiscale systems with large scale separation. 
}

\section{Introduction}
\label{sec:intro}
This chapter attempts a unified and comprehensive discussion of a number of studies that in about a decade have contributed to shape our nowadays's understanding of the implications, impacts and consequences for data assimilation when is applied to chaotic dynamics. The chapter presents a review of the essential results appeared in different studies, but for the first time together in a coherent treatment. In addition we present new original findings addressing two key aspects that were not covered in previous studies. The first treats the impact of data assimilation on a chaotic and multiscale system, the second concerns the consequences for nonlinear, non-Gaussian, data assimilation (a particle filter) in face of the chaotic nature of the underlying dynamics.

The exposition is organised as follows. We first discuss, in Sect.~\ref{sec:2}, the chaotic character of atmospheric and oceanic flows, and provide a treatment of the key mathematical concepts and tools that allow for characterising chaos and to ``measure'' the degree of instabilities. In doing so, we review classical results from dynamical system theory, including the {\it multiplicative ergodic theorem} and associated definition of Lyapunov (forward, backward and covariant) vectors and exponents. 

Section~3 analyses how a chaotic dynamics impacts data assimilation. Our focus is on Kalman filter (KF) and smoother (KS) in Sect.~\ref{sec:3.1}, and on their ensemble-based formulations, the ensemble Kalman filter (EnKF) and smoother (EnKS), in Sect.~\ref{sec:3.2}. 
Section~\ref{sec:3.1} contains primarily analytic results and treats KF and KS in linear systems, either purely deterministic (Sect.~\ref{sec:3.1.1}) or with stochastic additive noise (Sect.~\ref{sec:3.1.2}). Section~\ref{sec:3.2} is dedicated to nonlinear systems (deterministic, in Sect.~\ref{sec:3.2.1} and stochastic in Sect.~\ref{sec:3.2.2}) and on how the EnKF \citep{evensen2009} and the extended Kalman filter (EKF, \cite{ghil1991data}) works in this scenario. In Sect.~\ref{sec:3.2.1} we present original results on the impact of chaos on the performance of the EnKF in a coupled atmosphere-ocean model which possesses a degenerate-like spectrum of Lyapunov exponents, disentangling on the role of the quasi-null exponents.  

Section~\ref{sec:4} reverses the perspective and instead of studying the effect of chaos on data assimilation, reports on how properties of the dynamics have been used to devise adaptive observation strategies and ad-hoc data assimilation methods. In particular, in Sect.~\ref{sec:4.2}, we succinctly review the assimilation in the unstable subspace (AUS, \cite{palatella2013a}), a known approach that exploits the unstable-neutral subspace of the dynamics to perform the analysis. Remarkably, AUS was conceived before the findings reviewed in Sect.~\ref{sec:3}. The latter work was inspired by these early studies, and the need to provide mathematical rigor to clarify the mechanisms that made these early studies successful. In turn, the later work has furthermore provided the framework to generalize these early ideas to a variety of other types of dynamics.

Section~\ref{sec:5} presents what we consider two main areas of future developments.
In Sect.~\ref{sec:5} the paradigm of AUS is incorporated within a fully nonlinear data assimilation scheme, the particle filter \citep{van2019particle}. It is shown that observing in the directions of instabilities is effective but also that it is not deleterious to observe the stable directions. As opposed to the data or model sizes, our results suggest that the number of particles required to achieve good results scales with the size of the unstable-neutral subspace.  
Section~\ref{sec:5.2} discusses how the concept of random attractor can offer novel ways to handle the data assimilation problem in stochastic chaotic multi-scale systems with large scale separation. It also treats the implications of the numerical scheme on the output of the data assimilation cycle and ensemble-based forecasts, as well as poses some key questions for future studies. 

Final conclusions and a summary are drawn in Sect.~\ref{sec:6}.

\section{Chaos in atmospheric and oceanic flows}
\label{sec:2}

The atmosphere and the ocean are fluids that are described by the set of classical conservation laws of hydrodynamics, including the conservation of mass, momentum and energy \citep{vallis_2017}. For the atmosphere, these are often complemented by the conservation of moisture present in the air. These laws lead to a set of local dynamical equations describing the motion of each parcel of fluid. Given that these equations are nonlinear with complex interactions with the boundaries, realistic solutions cannot be obtained analytically and one must rely on numerical simulations starting from appropriate initial conditions.

Numerical simulations are based on discrete approximations in space and time of the dynamical equations, and are often accompanied with simplifications of the equations in order to describe appropriately the scales of interest. One of the most famous approximations is the geostrophic approximation which assumes a balance between the horizontal velocity field and the horizontal pressure gradient. Geostrophic balance is a good approximation for both the ocean and the atmosphere albeit at different spatial scales. The geostrophic approximation is at the basis of the models that will be used later in Sect.~\ref{sec:3.2.1}

Whatever the scale at which the atmospheric or the ocean fluids are observed, they display an apparently erratic evolution. This erratic behavior is also present in atmospheric, ocean and climate models when appropriate forcing are imposed. This feature should not be confused with randomness as most of the models used since the start of numerical modelling were purely deterministic. \cite{lorenz1963deterministic} showed in a simple low-order model that this erratic behavior is concomitant with the property of sensitivity to initial conditions, by which whatever small an error in the initial condition is, it will increase rapidly in time. This property implies that given the inevitable error in the initial conditions, any forecast will ultimately become useless as the error finally reaches an amplitude of the same order as the natural variability of the variable considered. The sensitivity to initial conditions and the consequent erratic-like evolution are the key properties of deterministic dynamical systems displaying {\it chaos}. 

This behaviour has been found in many atmospheric, oceanic and climate models \cite[see {\it e.g.,}][for a review]{vannitsem2017predictability}. 
In particular, a detailed investigation of the chaotic nature of the coupled ocean-atmosphere system that will be used in Sect.~\ref{sec:3.2.1} has been performed by \cite{vannitsemetal2015}.

\subsection{Measuring sensitivity to initial conditions}

The notion of sensitivity to initial conditions of dynamical solutions was already discovered and studied in a mathematical context by \cite{Poincare1899}. In the second half of the 20th century, this property was discovered in models of atmospheric and climate relevance \citep{Thompson1957,lorenz1963deterministic}. Its important practical implications drove the regain of interest in developing the appropriate mathematics in support of its description and understanding. These efforts culminated with the development of the ergodic theory of deterministic dynamical systems and chaos theory. In the following we shall briefly summarise what we consider to be the key developments that led to the definition of Lyapunov exponents and vectors. 

The importance of these mathematical objects stands on their ability to ``measure'' the degree of instabilities and thus to quantify the aforementioned sensitivity to initial conditions. While more rigorous mathematical treatments can be found in appropriate mathematical literature \citep[see {\it e.g.},][and references therein]{pikovsky2016lyapunov}, and we will invoke that rigour to a certain extent in the following sections, here we approach the discussion with a physical and intuitive angle.   
Further details can also be found in \cite{legras1996,barreira2002,Kuptsov2012}.

Let us write the evolution laws of a deterministic dynamical system in the form of a set of ordinary differential equations (ODEs),
\begin{equation}
\frac{\mathrm{d}\x}{\mathrm{d}t} = {\vec f}(\x, \boldsymbol{\sigma}),
\label{equat}
\end{equation}
where $\x$ is a vector containing the entire set of relevant variables $\x$ = $( x_1, ..., x_n)$ and $\bsigma$ represents a set of parameters. The discussion that follows holds also when system \eqref{equat} explicitly depends on time, thus being non-autonomous, provided it remains ergodic. 

Since the process of measurement is always subject to finite precision, the initial state is never known exactly. To study the evolution and the implications of such an error, let us consider an initial state displaced slightly from $\x_0$ by an initial error $\delta\x_0$.
The perturbed initial state, $\x_0+\delta\x_0$, generates a new trajectory in phase space and one can define the instantaneous error vector as the vector joining the
representative points of the reference trajectory and the perturbed one at a given time, $\delta \x(t)$. Provided that this perturbation is sufficiently small and smooth, its dynamics can be described by the linearized equation,
\begin{equation}
\frac{\mathrm{d}\delta \x}{\mathrm{d}t} \approx  \frac{\partial\vec F}{\partial\x}_{\vert \x(t)}
\delta\x ,
\label{linear}
\end{equation}
with a formal solution,
\begin{equation}
\delta\x(t) \approx \bM(t,\x(t_0)) \delta \x(t_0).
\end{equation}
The matrix $\bM$ is referred to as the ``fundamental matrix'' and is the resolvent of Eq.~\eqref{linear}, {\it i.e.} $\bM(t,\x(t_0))=e^{\int_{t_0}^{t}\frac{\partial\vec F}{\partial\x}_{\vert \x(t)}{\rm d}t}$. The fundamental matrix contains the information on the amplification of infinitesimally small perturbations.

In the context of the ergodic theory of deterministic dynamical systems, the Oseledets theorem \citep{Kuptsov2012} shows that 
the limit of the matrix
$(\mathbf{M}^\top \mathbf{M})^{1/[2(t-t_0)]}$, for time going to infinity, exists; let us refer to this limiting matrix as $\bf{S}$. The logarithm of its eigenvalues are called the {\it Lyapunov exponents} (LEs), whereas the full set of LEs is called the Lyapunov spectrum, and is usually represented in decreasing order. The
eigenvectors of $\bf{S}$, which are local properties of the flow (they change along the trajectory thus being time-dependent) and depend on the initial
time $t_0$, are called the {\it forward Lyapunov vectors} (FLVs) \citep{legras1996}.

The LEs are a powerful tool to ``measure'' chaos and to characterise the degree of instability of a system. For instance, LEs are averaged (asymptotic) indicators of exponential growth (LE>0) or decay (LE<0) of perturbations under the tangent-linear model. A deterministic chaotic system is uniquely characterised by having at least its leading LE larger than zero, {\it i.e.} LE$_1>0$. The sum of the LEs is equal to the average divergence of flow generated by Eq.~\eqref{equat} (see {\it e.g.} \cite{pikovsky2016lyapunov}). This means that in dissipative (conservative, {\it e.g.} Hamiltonian) systems the sum of the LEs is negative (zero): volumes in the phase space of a dissipative (conservative) dynamics reduce (are conserved) on average with time. Furthermore, autonomous continuous-in-time systems generally possess at least one LE=0, unless they converge to a motion-less state. This null exponent is related to perturbations aligned to the system's velocity vector ({\it i.e.} $\delta\x={\vec f}$), so although they shall fluctuate depending on the local flow, they will not on average decay nor growth. These last two properties can also be used as a check for numerical accuracy when computing the LEs.

The multiplicative ergodic theorem (MET) \citep[Theorem 2.1.2]{barreira2002} guarantees that, under general hypotheses, the eigenvalues of the matrix $(\mathbf{M}^\top \mathbf{M})^{1/[2(t-t_0)]}$ obtained for $t \rightarrow \infty$ are equivalent to the ones of the matrix $\mathbf{S'}=(\mathbf{M}\mathbf{M}^\top)^{1/[2(t-t_0)]}$ when $t_0 \rightarrow -\infty$.  The equivalence of the spectrum of  $\mathbf{S}$ and $\mathbf{S'}$  is not generically true for the fundamental matrix of an arbitrary, linear dynamical system; however, the MET guarantees that this is a fairly generic property of the resolvent of the \emph{tangent-linear model of a nonlinear dynamical system}. If the flow of the time derivative ${\vec f}$ is a $\mathcal{C}^1$  diffeomorphism of a compact, smooth, Riemannian manifold $M$, then the MET assures that the LEs are defined equivalently by the log-eigenvalues of $\mathbf{S}$ or $\mathbf{S'}$ for any initial condition $\mathbf{x}(t_0)$ and that the LEs are unique on a subset of $M$ of full measure with respect to any ergodic invariant measure $\mu$ of the flow. Note that, without the condition of ergodicity of $\mu$, the LEs may be well-defined point-wise, but the specific values and their multiplicity may depend on the initial condition $\mathbf{x}(t_0)$.

The matrices $\mathbf{S}$ and $\mathbf{S}'$ are symmetric.  However, contrary to the eigenvalues, the eigenvectors of these two matrices are not equivalent due to the asymmetric character of the fundamental matrix $\bM$ in forward- and reverse-time. The eigenvectors of the latter are called the {\it backward Lyapunov vectors} (BLVs). Theoretically, each matrix can be evaluated at the same place along the reference trajectory $\x(t)$ and their orthogonal eigenvectors can be computed as $\mathbf{L}^{\mathrm{f},i}_{t}$ and
$\mathbf{L}^{\mathrm{b},i}_{t}$ for $\mathbf{S}$ and $\mathbf{S}'$, respectively, where it is understood that the time-dependence on $t$ is with respect to the linearization of the dynamics at $\vec{x}(t)$. There exist \emph{Oseledet} subspaces $W^i_{t}$,
\begin{equation}
W^i_{t} = \mathbf{L}^{\mathrm{b},1}_{t}\oplus ... \oplus \mathbf{L}^{\mathrm{b},i}_{t} \cap \mathbf{L}^{\mathrm{f},i}_{t} \oplus ... \oplus  \mathbf{L}^{\mathrm{f},N}_{t} ,
\label{wi}
\end{equation}
with $\oplus$ being the direct product \citep{Ruelle1979}, that have the important properties of being invariant under the effect of the fundamental matrix, such that \begin{equation}
{\bf M} (\tau ,\vec x(t)) W^i_{t} = W^i_\tau.
\end{equation}
Due to their orthogonal nature, the FLVs and the BLVs require by definition the choice of a norm and of an inner product. Nevertheless, the Oseledet subspaces themselves do not have this dependence; in this way, they can be considered to embed more invariant information about the dynamics. The decomposition of the tangent-linear space into these covariant subspaces is commonly known as the ``Oseledet splitting'' or decomposition. The classical form of the MET thus guarantees that the Oseledet splitting is well-defined and consistent with probability one over all initial conditions of the attractor, with respect to the invariant, ergodic measure. Other covariant splittings of the tangent-linear model, such as by exponential dichotomy, exist under more general forms of the MET \citep{froyland2013computing}.

When the Lyapunov spectrum is non-degenerate, one can define a time-varying basis, subordinate to the Oseledet spaces, $W^i_{t} = \mathrm{span}\left\{\mathbf{L}^{\mathrm{c},i}_{t}\right\}$, such that
\begin{equation}
{\bf M} (\tau ,\vec x(t)) \mathbf{L}^{\mathrm{c},i}_{t} = \alpha_i (\tau, \vec x(t)) \mathbf{L}^{\mathrm{c},i}_{\tau},
\label{ampli}
\end{equation}
where $\parallel \mathbf{L}^{\mathrm{c},i}_{t} \parallel = \parallel \mathbf{L}^{\mathrm{c},i}_\tau \parallel $, and $\alpha_i (\tau, \vec x(t))\in\mathbb{R}$ 
describes an amplification factor. The vectors $\mathbf{L}^{\mathrm{c},i}_{t}$ are known as the {\it covariant Lyapunov vectors} (CLVs). 
In the long-time limit, the amplifications $\alpha_i(\tau, \vec{x}(t))$ can be associated
to the LEs as,
\begin{equation}
\pm \lambda_i =  \lim_{\tau \rightarrow \pm \infty} \frac{1}{\tau} ln \,\, \vert\alpha_i (\tau, \vec x(t))\vert=
 \lim_{\tau \rightarrow \pm \infty} \lambda_i^\tau(\x(t))
 \label{eq:lle_limit}
\end{equation}
where we indicate by $\lambda_i^\tau(\x(t)) $ the average of the growth rate taken over a time window $\tau$ starting at $t$ at position $\x(t))$, and are commonly known as the local Lyapunov exponents (LLEs) \citep[see chapter 5]{pikovsky2016lyapunov}.

Throughout the chapter we will denote the matrix with columns corresponding to the full tangent linear space ordered basis of BLVs/FLVs/CLVs at time $t$ as $\mathbf{L}^{\mathrm{e}}_{t}$ for $\mathrm{e}\in\{\mathrm{b},\mathrm{f},\mathrm{c}\}$ respectively. A sub-slice of this matrix of Lyapunov vectors corresponding, inclusively, to columns $i$ through $j$ will be denoted $\mathbf{L}^{\mathrm{e}, i:j}_{t}$.  In the following, the relationships between these Lyapunov vectors and their asymptotic dynamics will be key to understanding the predictability of chaotic systems. For this purpose, we will largely use the BLVs and the CLVs, and to a lesser extent the FLVs.

Note that the basis $\left\{\mathbf{L}^{\mathrm{c},i}_{t}\right\}_{i=1}^{N_x}$ is not orthogonal in general and, in fact, the angles between any two Oseledec subspaces $W_i$ and $W_j$ are not in general bounded away from zero in their limits in forward- or reverse-time. For this reason, coordinate transformations to covariant Oseledet bases are not generally numerically well-conditioned asymptotically, and therefore hard to compute. The property of \emph{integral separation} \citep{dieci2002lyapunov}, describing the stability of the Lyapunov spectrum under bounded perturbations of the tangent-linear equations, ensures that the angles between the covariant subspaces will remain bounded away from zero, but this is a strong condition and it is not as generic of a property as the existence of the Oseledet decomposition under the MET.  However, if a dynamical system is integrally separated as above, there exists a well-defined, numerically well-conditioned transformation of coordinates of the tangent-linear model for which the action of the resolvent $\mathbf{M}$ can be expressed as a block-diagonal matrix with each block describing the invariant dynamics of a single Oseledet space. For degenerate spectrum, these blocks may be upper-triangular, but for non-degenerate spectrum this representation of the resovlent becomes a strictly diagonal matrix; see Theorem 5.4.9 of \cite{adrianova1995introduction} for the classical result, or Theorem 5.1 of \cite{dieci2007lyapunov} and \cite{froyland2013computing} for more recent extensions.

\section{Data assimilation in chaotic systems - How the dynamics impacts the way we assimilate data}
\label{sec:3}

The high sensitivity of a chaotic dynamical system to the initial condition makes it hard to forecast it, even when their evolution equations are perfectly known. Indeed the typical error grows exponentially over time with a rate given by the largest positive LE. With a view to forecasting, one has no choice but to regularly correct its trajectory using information on the system state obtained through observations. This is the primary goal of data assimilation (DA) and has been key to the success of numerical weather forecasting.
We refer to \citet{kalnay2003atmospheric,asch2016data} and references therein for textbooks and reviews on geophysical DA. 

When framed in a Bayesian formalism, the goal of DA is to estimate the conditional probability density function (pdf) of the system state knowing observations of that system. For high-dimensional systems, only approximations of this pdf can be obtained, among which the Gaussian approximation is the most practical and common \citep{carrassi2018data}.
In the case of Gaussian statistics of the error and linear dynamics, the conditional pdf can be obtained analytically: it is Gaussian, and can be sequentially computed using the Kalman filter (KF) \citep{kalman1960}. Directly inspired from the Kalman filter, the extended Kalman filter (EKF) offers an approximate DA scheme when the operators are nonlinear \citep{ghil1991data}. However, it can hardly be used in the context of high-dimensional systems, where it has to be replaced with the ensemble Kalman filter (EnKF) \citep{evensen2009data}. In the EnKF, the covariance matrices are represented by state perturbations which are representative of the errors and, together with the state mean, form a limited-size ensemble of state vectors.

Understanding how the ensemble of the EnKF evolves under the action of the DA scheme and the forecast model dynamics is important.
Several numerical results suggest that the skills of ensemble-based DA methods
in chaotic systems are related to the instabilities of the underlying dynamics \citep{ng2011role}.
Numerical evidences exist that some asymptotic properties of the ensemble-based covariance matrices (rank, span, range)
relate to the unstable modes of the dynamics \citep{sakov2008deterministic, carrassi2009}.
Nevertheless, a better, more profound, understanding of these results was needed to aim at designing reduced rank, computationally cheap, formulations of the filters. 

Analytic results that have shed lights on the behaviour of filters and smoothers on chaotic dynamics, and explained the numerically observed properties, have been obtained for linear dynamics. Section~\ref{sec:3.1.1} reviews those findings in the case of deterministic systems without model error, based on the work by \cite{gurumoorthy2017} and \cite{bocquet2017degenerate}. Section~\ref{sec:3.1.2} treats their extension to the case of stochastic systems, following the work by \cite{grudzien2018asymptotic,grudzien2018chaotic}. 
In the case of nonlinear dynamics, robust numerical evidences will be described by the original experiments in Sect.~\ref{sec:3.2}, that extent the previous findings by \cite{bocquet2017four}. 
    \subsection{Linear dynamics: the effect of chaos on the Kalman filter and smoother}
\label{sec:3.1}
        \subsubsection{Perfect and deterministic dynamics}
\label{sec:3.1.1}

At time $t_k$, let $\bx_k \in {\mathbb R}^\Nx$ and $\by_k \in {\mathbb R}^\Ny$ be the state and observation vector, respectively.
Let us assume linear evolution model dynamics $\bM_k$ and observation model $\bH_k$ such that
\begin{subequations}
\begin{align}
  \label{eq:dynmodel}
  \bx_{k} &= \bM_{k}\bx_{k-1} + \bw_k , \\
  \label{eq:obsmodel}
  \by_{k} &= \bH_k \bx_k + \bv_k .
 \end{align}
 \end{subequations}
The model and observation noises, $\bw_k$ and $\bv_k$, are assumed mutually independent, zero-mean Gaussian white
sequences with statistics
\be
   {\rm E}[\bv_k\bv_l\T] = \delta_{k,l}\bR_k , \quad  {\rm E}[\bw_k\bw_l\T] = \delta_{k,l}\bQ_k , \quad {\rm E}[\bv_k\bw_l\T] = \bzero \, .
\ee
In the Kalman filter, which yields the optimal DA solution with such assumptions, the forecast error covariance matrix $\bP_k$ satisfies the recurrence
\be
\label{eq:recurrence}
  \bP_{k+1} = \bM_{k+1}\(\bI+\bP_k \bOmega_k\)^{-1}\bP_k\bM_{k+1}\T +\bQ_{k+1}, \,
\ee  
where 
\be 
\bOmega_k \equiv \bH_k\T\bR_k^{-1}\bH_k ,
\ee
is the {\it precision matrix} of the observations mapped into state space.
In the absence of model error, {\it i.e.} $\bQ_k \equiv \bzero$, \citet{gurumoorthy2017} proved rigorously that in the full-rank KF (in particular $\bP_0$ is full rank), $\bP_k$ collapses onto the unstable-neutral subspace.

We shall summarise in the following some key results that apply to the full-rank but also to the degenerate case, {\it i.e.} even if $\bP_0$ is of arbitrary rank and the initial errors of the DA scheme only lie in a subspace of ${\mathbb R}^\Nx$. In particular they apply to the EnKF if the dynamical and observations operators are both linear. We recall them here, but readers can find full details in \citet{bocquet2017degenerate, bocquet2017four}.

\paragraph*{{\bf Result 1}: \underline{Bound on the free forecast error covariance matrix}}

Let us define the resolvent of the dynamics from $t_l$ to $t_k$ as $\bM_{k:l}=\bM_k \bM_{k-1} \cdots \bM_{l+1}$, with the convention that $\bM_{k,k}=\bI$.
The first key result is the following inequality in the set of the semi-definite symmetric matrices:
\be
\label{eq:bound0}
\bP_k \le \bM_{k:0}\bP_0 \bM_{k:0}\T + \bXi_k \eqqcolon \bP^{{\rm free}}_k,
\ee
where
\be
\bXi_0 \equiv \bzero \, \quad \text{and for} \, k\ge 1 \quad \bXi_k \equiv
\sum_{l=1}^k \bM_{k:l}\bQ_l \bM_{k:l}\T ,
\label{eq:controllability_matrix}
\ee
is known as the \emph{controllability} matrix \citep{jazwinski1970}, and the suffix ``free'' is used to refer to the forecast error of the system unconstrained by data.
In the absence of model noise ($\bQ_k\equiv \bzero$ for the rest of this section), it reads
\be
\label{eq:bound1}
  \bP_k \le \bM_{k:0}\bP_0 \bM_{k:0}\T .
\ee
Assuming the dynamics to be non-singular, the column subspace of the forecast error covariance matrix satisfies
\be
  \im(\bP_k) = \bM_{k:0}\(\im(\bP_0)\) .
\ee
If $n_0$ is the dimension of the unstable-neutral subspace of the dynamics, it can further be shown that
\be
\label{eq:minrank}
\lim\limits_{k \rightarrow \infty}\rank(\bP_k) \le \min\left\{ \rank(\bP_0), n_0 \right\} ,
\ee
which is a first proof of the collapse of the error covariance matrix (actually its column space) onto the unstable and neutral subspace of the dynamics.

\paragraph*{{\bf Result 2}: \underline{Collapse onto the unstable subspace}}

Let $\sigma_i^k$, for $i = 1, \dots, \Nx$ denote the eigenvalues of $\bP_k$ ordered as $\sigma_1^k \ge \sigma_2^k \dots \ge \sigma_\Nx^k$. It was shown that
\be
\label{eq:convrate}
\sigma_{i}^k \leq \beta_i \exp\(2\lambda^k_{i} k \),
\ee
for some $\beta_i > 0$, where $\lambda^k_{i}$ is a log-singular value of $\bM_{k:0}$ that converges to the LE $\lambda_{i}$.
This gives an upper bound for all eigenvalues of $\bP_k$ and a rate of convergence for the $\Nx-n_0$ smallest ones.
Moreover, if $\bP_k$ is uniformly bounded, it can further be shown that
the stable subspace of the dynamics is asymptotically in the null space of $\bP_k$, {\it i.e.},
\be
\lim_{k \rightarrow \infty} \left\| \bP_k \mathbf{L}^{\mathrm{b},i}_k \right\| = 0
\ee
 for all $i>n_0$; this extends to any norm and linear combination of these vectors.

\paragraph*{{\bf Result 3}: \underline{Explicit dependence of $\bP_k$ on $\bP_0$}}

To study the dependence of $\bP_k$ on $\bP_0$, it has been shown that the forecast error covariance matrix can be written as
\be
\label{eq:final2}
  \bP_k = \bM_{k:0}\bP_0\bM_{k:0}\T \(\bI + \bGamma_k \bM_{k:0}\bP_0\bM_{k:0}\T\)^{-1},
\, 
\ee
where the matrix
\be
\bGamma_k \equiv \sum_{l=0}^{k-1} \bM_{k:l}^{-\top} \bOmega_l \bM_{k:l}^{-1} ,
\ee
is known as the \emph{information matrix} and it measures the \emph{observability} of the system by propagating the precision matrices $\bOmega_l$ up to $t_k$. 

An alternative formulation of Eq.~\eqref{eq:final2} is
\be
\label{eq:final}
  \bP_k = \bM_{k:0}\bP_0  \left[ \bI + \bTheta_k\bP_0 \right]^{-1}\bM_{k:0}\T ,
\ee
where 
\be
\label{eq:thetadef}
\bTheta_k \equiv \bM_{k:0}\T \bGamma_k \bM_{k:0} = \sum_{l=0}^{k-1} \bM_{l:0}^{\top} \bOmega_l \bM_{l:0},
\ee
is also an \emph{information} matrix, directly related to the \emph{observability} of the DA system, but pulled back at the initial time $t_0$.
Equation \eqref{eq:final} is of key importance because it allows to study the asymptotic behaviour of $\bP_k$, {\it i.e.} the filter ``believed'' error,
using the asymptotic properties of the
dynamics.

\paragraph*{{\bf Result 4}: \underline{Asymptotics of $\bP_k$}}

For $\bP_k$ to forget about $\bP_0$ as $k$ tends to infinity, it was shown that one can impose the following sufficient conditions:

\begin{itemize}
    \item {\bf Condition 1:}
Recall that the FLVs at $t_0$
associated to the unstable and neutral exponents are the columns of $\mathbf{L}^{\mathrm{f},1:n_0}_{0} \in {\mathbb R}^{\Nx \times n_0}$. Moreover, let define the anomaly matrix $\bX\in\mathbb{R}^{\Nx \times n_0}$ such that $\bP=\bX\bX^{\rm T}$. 
The condition reads
\be
\label{eq:condition1}
\rank \( \bX_0\T \mathbf{L}^{\mathrm{f},1:n_0}_{0} \) = n_0 .
\ee
In practice the initial ensemble anomalies $\bX_0$ projects onto the first $n_0$ FLVs at $t_0$.

\item {\bf Condition 2:}
The model is sufficiently observed so that the unstable and neutral directions
remain under control, {\it i.e.}, there exits  $\varepsilon > 0$ such that

\be
\label{eq:condition2}
\left(\mathbf{L}^{\mathrm{b},1:n_0}_{k}\right)\T \bGamma_k \mathbf{L}^{\mathrm{b},1:n_0}_{k} > \varepsilon \bI.
\ee

\item {\bf Condition 3:}
Furthermore, for any neutral BLV we have
\be
\label{eq:condition3}
\lim_{k \rightarrow \infty} 
\left(\mathbf{L}^{\mathrm{b},n_0}_{k}\right)\T \bGamma_k \mathbf{L}^{\mathrm{b},n_0}_{k} = \infty ,
\ee
implying that the neutral direction is well observed and controlled. 
\end{itemize}

Under these three conditions, we obtain
\be
\lim_{k \rightarrow \infty} \left\{ \bP_k -
\mathbf{L}^{\mathrm{b},1:n_0}_{k}\left[ \left(\mathbf{L}^{\mathrm{b},1:n_0}_{k}\right)\T \bGamma_k \mathbf{L}^{\mathrm{b},1:n_0}_{k}\right]^{-1}\left(\mathbf{L}^{\mathrm{b},1:n_0}_{k}\right)\T \right\} = \bzero .
\ee
Hence, the asymptotic sequence does not depend on $\bP_0$, but only $\bGamma_k$,
{\it i.e.} the dynamics.
It can also be shown that the neutral modes have a peculiar role and a long lasting influence: their influence on the current estimate decreases sub-exponentially.

\paragraph*{{\bf Result 5}: \underline{From the degenerate KF to the square-root EnKF and EnKS}}

The recurrence Eq.~\eqref{eq:final} can be reformulated in a factorised form which is suited to the 
square-root EnKF. The standard perturbation decomposition of the forecast error covariance is  
\be
\bP_k = \bX_k\bX_k\T ,
\ee
where $\bX_k$ is the matrix of centred perturbations (the anomaly matrix aforementioned).
A right-transform update formula can then be obtained from \eqref{eq:final}:
\be
\label{eq:main-filter-sqrt}
\bX_k = \bM_{k:0}\bX_0  \left[ \bI + \bX_0\T\bTheta_k\bX_0 \right]^{-1/2}\bUpsilon_k ,
\ee
where $\bUpsilon_k$ is an arbitrary orthogonal matrix such that $\bUpsilon_k \bone = \bone$, where $\bone$ is the vector $[1 \ldots 1]\T$ defined in the ensemble subspace.
It is equivalent to the left-transform update formula
\be
\label{eq:main-filter-sqrt-bis}
\bX_k = \left[ \bI + \bM_{k:0}\bP_0 \bM_{k:0}\T \bGamma_k \right]^{-1/2}\bM_{k:0}\bX_0 \bUpsilon_k .
\ee
The importance of Eqs.~\eqref{eq:main-filter-sqrt} and \eqref{eq:main-filter-sqrt-bis} stands on the fact that with linear models, Gaussian observation and initial errors, the (square-root) degenerate KF (with $\bX\in\mathbb{R}^{N_x\times n}$ and $n<N_x$) is equivalent to
the square-root EnKF and can serve as a proxy to the EnKF applied to nonlinear models.

All of these results can be generalised to linear smoothers.
In particular, the smoother forecast error covariance matrix is similar to that of the filter, {\it i.e.} given by \eqref{eq:final2} (or \eqref{eq:final}) but with the following modified information matrix:
\be
\widehat{\bGamma}_k = \bGamma_k + \sum_{l=k}^{k+L-S} \bM_{k:l}^{-\mathrm{T}}\bOmega_k \bM_{k:l}^{-1},
\ee
where $L$ is the lag of the smoother (how far in the past observations are accounted for) and $S$ tells by how many time steps the smoother's window is shifted between two consecutive updates. Note that $\widehat{\bGamma}_k \ge \bGamma_k$ (using the Loewner order on the set of semi-definite positive matrices) reflecting the general higher amount of information incorporated within a smoother update relative to a filter. Therefore the asymptotic sequences for the filter (right-hand side) and smoother (left-hand-side) follow the inequality:
\begin{align}
\begin{split}
&\mathbf{L}^{\mathrm{b},1:n_0}_{k}\left[ \left(\mathbf{L}^{\mathrm{b},1:n_0}_{k}\right)\T \widehat{\bGamma}_k \mathbf{L}^{\mathrm{b},1:n_0}_{k}\right]^{-1}\left(\mathbf{L}^{\mathrm{b},1:n_0}_{k}\right)\T  \le \\ 
&\mathbf{L}^{\mathrm{b},1:n_0}_{k} \left[ \left(\mathbf{L}^{\mathrm{b},1:n_0}_{k}\right)\T \bGamma_k \mathbf{L}^{\mathrm{b},1:n_0}_{k}\right]^{-1}\left(\mathbf{L}^{\mathrm{b},1:n_0}_{k}\right)\T .
\end{split}
\end{align}

The linear smoothers can serve as a proxy to the ensemble Kalman smoother \citep{evensen2009data} or the iterative ensemble Kalman smoother (IEnKS) \citep{bocquet2014} applied to nonlinear models \citep{bocquet2017four}. With the IEnKS, where the true state trajectory is even better estimated and the errors are reduced, the collapse of the perturbations onto the unstable and neutral subspace is expected to be even faster.

        \subsubsection{Stochastic dynamics}
\label{sec:3.1.2}
The above perfect-deterministic model configuration demonstrates how chaos shapes the inferences of the posterior.  Asymptotic statistics are determined by the ability of the filter to control the growth of initial error in the unstable-neutral subspace with respect to its sensitivity to observations therein.  However, in realistic DA, additional forecast errors are introduced throughout the forecast cycle due to the inadequacy of numerical models in representing reality.  One classical approach to treat these model errors is to represent them as additive or multiplicative noise \citep{jazwinski1970}.  Oseledet's theorem and the Lyapunov spectrum are also formulated for such systems of stochastic differential equations (SDEs) and discrete maps.   

Suppose $\{\mathbf{v}^i\}_{i=0}^d$ is a collection of $\mathcal{C}^3$ vector fields on $M \subset \mathbb{R}^{N_x}$, a smooth manifold without boundary. Define $\{W_t^i \}^d_{i=1}$ where each $W_t^i$ is an independent Wiener process defined on the probability space $\left(\Omega,\mathcal{F},P\right)$.  This describes a generic Stratonovich SDE,
\begin{align}
\frac{\mathrm{d}\mathbf{x}}{\mathrm{d} t} = \mathbf{f}(\mathbf{x}, \boldsymbol{\sigma}, \omega) \triangleq \mathbf{v}^0\left(\mathbf{x}, \boldsymbol{\sigma} \right) + \sum_{i=1}^d \mathbf{v}^i\left(\mathbf{x}, \boldsymbol{\sigma}\right)\circ W^i_t(\omega).
\label{eq:strat}
\end{align}
For fixed $\boldsymbol{\sigma}$, if $\{\mathbf{v}^i\}_{i=1}^d$ span the tangent space $T_{\mathbf{x}} M$ for each $\mathbf{x}\in M$, then the system of SDEs gives rise to a unique probability measure $\mu$ on $M$ that is invariant with respect to the random flow induced by the system of SDEs.
For $\mu \times P$ almost every $(\mathbf{x}, \omega)\in M \times \Omega$, the Lyapunov exponents and their multiplicity are defined and only depend on $\mathbf{x}$ \citep{liu2006smooth}[theorem 2.1].  For the SDE, $\mathbf{f}(\mathbf{x}, \boldsymbol{\sigma},\omega)$, the tangent linear model is once again defined as in Eq. (\ref{linear}) from which the exponents and vectors can be computed as usual \citep{pikovsky2016lyapunov}[chapters 2 and 8].

The analysis from perfect-deterministic models thus extends to stochastically forced models as in Eq. \ref{eq:dynmodel}, but with key differences.
 Firstly, the forecast error covariance can generally be considered to be of full rank due to the injection of the stochastic forcing $\mathbf{w}_k$ into arbitrary subspaces;  the standard controllability assumption actually guarantees that it is of full rank after a sufficient lead time \citep{jazwinski1970}[lemma 7.3].  Stochastic perturbations $\mathbf{w}_k$ are, moreover, subject to growth and decay rates of the LLEs, $\lambda_i^\tau$ of Eq. (\ref{eq:lle_limit}).  These LLEs are distributed about the LE such that even asymptotically stable modes may have transient periods of rapid growth.  
 
 To make the analysis tractable, assume that the dynamics are stationary in the sense that the recursive QR algorithm \citep{shimada1979, benettin1980} converges uniformly to the theoretical LEs uniformly in number of iterations from any initial time. While the forecast error covariance is not generally reduced-rank, the stable dynamics is actually sufficient to uniformly bound the forecast error variances in the stable subspace \emph{without any assimilation}.

\paragraph*{{\bf Result 6}: \underline{Uniform bounds on the variance of errors in the stable subspace}}

Recall that $\mathbf{L}^{\mathrm{b}, i}_k$ is the $i$-th BLV and that $\mathbf{P}^\mathrm{free}_k$ is the free forecast error covariance ({\it e.g.} without DA).  For $\lambda_i < 0 $, define $\epsilon>0$ such that $\exp\left\{2\lambda_i + \epsilon\right\} < 1$.  By the stationarity assumed above, there exists $N_{\epsilon}$ such that if $k-l > N_{\epsilon}$, then
\begin{align}
- \epsilon < \frac{1}{k-l} \log\left( \parallel \mathbf{M}_{k:l}^\top \mathbf{L}^{\mathrm{b},i}_k \parallel\right) - \lambda_i  < \epsilon.
\end{align}
Assuming that the system is uniformly completely controllable and
$\mathbf{Q}_k \leq q_\mathrm{sup} \mathbf{I}_n < \infty$ for all $k$, the controllability matrix, Eq. (\ref{eq:controllability_matrix}), is uniformly bounded above by $C_{N_{\epsilon}}\mathbf{I}_n$.  A variation on Eq. (\ref{eq:bound0}) can be used to obtain a uniform bound on the $i$-th variance of $\mathbf{P}^\mathrm{free}_k$ in the basis of the BLVs as
\begin{align}
\limsup_{k\rightarrow \infty} \left(\mathbf{L}^{\mathrm{b},i}_k\right)^\top \mathbf{P}_k^\mathrm{free} \mathbf{L}^{\mathrm{b},i}_k \leq C_{N_{\epsilon}} + \frac{q_{\sup} \exp\left\{2(\lambda_i +\epsilon)N_{\epsilon} +1\right\}}{1-\exp\left\{2(\lambda_i+\epsilon)\right\}}.
\end{align}

This bound represents the competing forces of the transient growth rates of recently introduced perturbations in the controllability matrix, and perturbations that adhere to their aysmptotic log-average rate of decay in the stable subspace within a margin of $\epsilon$ \citep{grudzien2018asymptotic}[proposition 2 and corollary 3]. This demonstrates that, if assimilation prevents error growth in the span of the unstable-neutral BLVs, errors in the span of the stable BLVs can be neglected without relinquishing filter boundedness.  However, this does not state whether the error in the span of the stable subspace will remain within tolerable bounds.  

Redefine $q_{\sup}$ such that $\mathbf{P}_0^\mathrm{free}, \mathbf{Q}_k \leq q_\mathrm{sup}\mathbf{I}_n$ for all $k$.  The variance of the free forecast in the $i$-th BLV is bounded directly as
\begin{align}\begin{split}
 \left(\mathbf{L}^{\mathrm{b},i}_k\right)^\top \mathbf{P}^\mathrm{free}_k \mathbf{L}^{\mathrm{b},i}_k  & = \left(\mathbf{L}^{\mathrm{b},i}_k\right)^\top\mathbf{M}_{k:0}\mathbf{P}_0^\mathrm{free} \mathbf{M}_{k:0}^\top \mathbf{L}^{\mathrm{b},i}_k +  \sum_{l=1}^k \left(\mathbf{L}^{\mathrm{b},i}_k\right)^\top \mathbf{M}_{k:l} \mathbf{Q}_l \mathbf{M}^\top_{k:l}  \mathbf{L}^{\mathrm{b},i}_k \\
& \leq  q_\mathrm{sup}\sum_{l=0}^k \left(\mathbf{L}^{\mathrm{b},i}_k\right)^\top \mathbf{M}_{k:l} \mathbf{M}^\top_{k:l}  \mathbf{L}^{\mathrm{b},i}_k = q_\mathrm{sup} \sum_{l=0}^k \parallel \left(\mathbf{T}^\top_{k:l}\right)^{i} \parallel^2 ,
\end{split}
\label{eq:freeforecast}
\end{align}
where $\parallel \left(\mathbf{T}^\top_{k:l}\right)^{i} \parallel$ is the norm of the $i$-th row of $\mathbf{T}_{k:l}$ in the recursive QR decomposition of the propagator, $\mathbf{M}_{k:l}= \mathbf{L}^{\mathrm{b}}_k \mathbf{T}_{k:l}\left(\mathbf{L}^{\mathrm{b}}_l\right)^\top$.  
The sum 
\begin{align}
\Psi_k^{i} \triangleq\sum_{l=0}^k \parallel \left(\mathbf{T}^\top_{k:l}\right)^{i} \parallel^2
\label{eq:psi_i}
\end{align}
describes the invariant evolution for the $i$-th variance of the free forecast error covariance matrix in the basis of BLVs.  In the case that $\mathbf{Q}_k=q\mathbf{I}_n$ for all $k$, Eq. (\ref{eq:freeforecast}) becomes an equality and $\Psi_k^{i}$ can be interpreted as the evolution of the $i$-th variance when $q=1$.  As $k-l$ grows $\parallel \left(\mathbf{T}_{k:l}^\top\right)^i\parallel$ converges exponentially to zero while for $k-l$ close to one this describes transient dynamics in the basis of the BLVs. Although $\Psi^i_k$ is guaranteed to be uniformly bounded in $k$ \citep{grudzien2018asymptotic}[corollary 3], numerical simulations demonstrate how this uniform bound can be extremely large.
 
\begin{figure}[ht]
\center
\includegraphics[width=.85\linewidth]{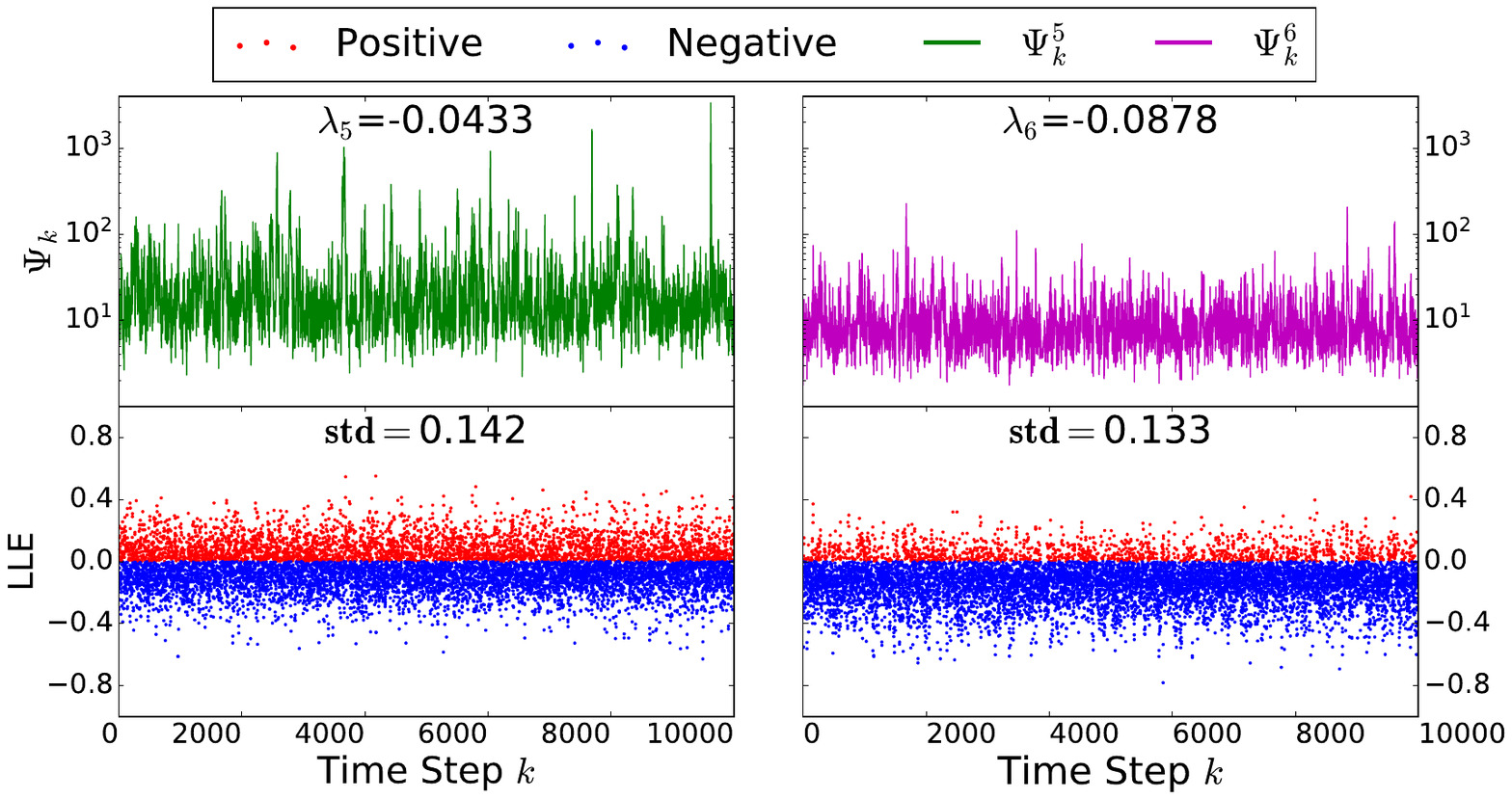}
\caption{\textbf{Upper:} time series of $\Psi^{5}_k$ and $\Psi^{6}_k$ as defined in Eq. (\ref{eq:psi_i}).  \textbf{Lower:} LLEs of $\mathbf{L}^{\mathrm{b},5}$ and $\mathbf{L}^{\mathrm{b},6}$. Adapted from \cite{grudzien2018asymptotic}}
\label{fig:bnd_v_lle}
\end{figure}

Figure~\ref{fig:bnd_v_lle}, presents an example from \cite{grudzien2018asymptotic} of the free forecast error variance, $\Psi_k^i$, over $10^4$ forecast cycles where the model propagator $\mathbf{M}_k$ is defined by the evolution of the tangent linear model of the Lorenz-96 system \citep{lorenz96} in $N_x=10$ dimensions, with an interval between observations of $0.1$, and $\mathbf{Q}_k \triangleq \mathbf{I}_{N_x}$. Although this model is generated from the underlying nonlinear Lorenz system, the state model for the experiment is treated as a discrete linear model using only the tangen-linear resolvent as above. This model has three unstable, one neutral and six stable LEs. While $\lambda_5$ and $\lambda_6$ are negative, $\mathbf{L}^{\mathrm{b},5}_k$ and $\mathbf{L}^{\mathrm{b},6}_k$ experience frequent transient instabilities in the timeseries of their LLEs (see bottom row in Fig.~\ref{fig:bnd_v_lle}). The LLEs of $\mathbf{L}^{\mathrm{b},5}_k$ have more intense growth, reflected in the differences between $\Psi^5_k$ and $\Psi^6_k$ (top row): the maximum of $\Psi_k^6$ is on the order of $\mathcal{O}(10^{2})$ and the mean is approximately 28; for $\Psi_k^5$ the max is of $\mathcal{O}\left(10^3\right)$ and the mean is approximately 808. This suggests that, as opposed to the case of deterministic dynamics, for successful DA in stochastic systems with additive model error, it is necessary to control the growth of forecast errors also in the span of weakly stable BLVs. While errors in this span will not grow indefinitely and remain bounded, if those directions are left uncontrolled their error bounds can be practically too large for any meaningful state estimation purposes. 

\paragraph*{{\bf Result 7}: \underline{The unfiltered-to-filtered error upwell and the need for inflation}}

Motivated by the results above, suppose that an approximate, reduced-rank Kalman estimator is defined such that the resulting forecast error covariance and the Kalman gain have image (column) spaces constrained to the span of the leading $r\geq n_0$ BLVs, $\mathbf{L}^{\mathrm{b}, 1:r}_k$. For perfect and deterministic dynamics, this estimator is asymptotically equivalent to the optimal Kalman filter by the results of Sect.~\ref{sec:3.1.1}. On the other hand, Result 6 establishes that, in the presence of additive model errors, the variance of the unfiltered error in the span of the trailing $N_x-n_0$ BLVs will remain finite, albeit bounded as per Eq.~\eqref{eq:freeforecast}. When $r>n_0$ this reduced-rank Kalman filter will correct the $r-n_0$ stable modes in addition to unstable-neutral modes, thereby reducing the variance of the free forecast errors below the bounds pictured in Fig.~\eqref{fig:bnd_v_lle}.

\cite{grudzien2018chaotic} derive the full forecast error covariance dynamics for the reduced-rank Kalman estimator described above. Note that, as opposed to the free forecast error covariance described in relation to Result 6, the discussion pertains now to the forecast error covariance cycled within the KF. 
While the standard reduced-rank KF formalism would write the recursion for the forecast error covariance entirely within the span of $\mathbf{L}^{\mathrm{b},1:r}$, it is proven in \cite{grudzien2018chaotic} that forecast errors in the column span of trailing $\mathbf{L}^{\mathrm{b},r+1:N_x}_k$ (those left ``uncorrected'' by DA) are transmitted into the column span of $\mathbf{L}^{\mathrm{b}, 1:r}_k$. This ``error upwell'' is a consequence of the KF rank-reduction within the first $r$ BLVs and is driven by the upper triangular dynamics of the BLVs in the recursive QR algorithm.  
Therefore, neglecting the contribution of the ``upwelling'' of error from the trailing to the leading BLVs, as in the standard recursion, leads to a systematic underestimation of the true forecast error in the presence of additive noise. Furthermore, because the leading $r$ BLVs share the same span as the leading $r$ Oseledet spaces, Eq.~\eqref{wi}, the upwelling of errors from the span of the trailing BLVs to the leading BLVs holds for any estimator that is restricted to the span of the leading $r$ covariant subspaces.

\begin{figure}[t]
\center
\includegraphics[width=.75\linewidth]{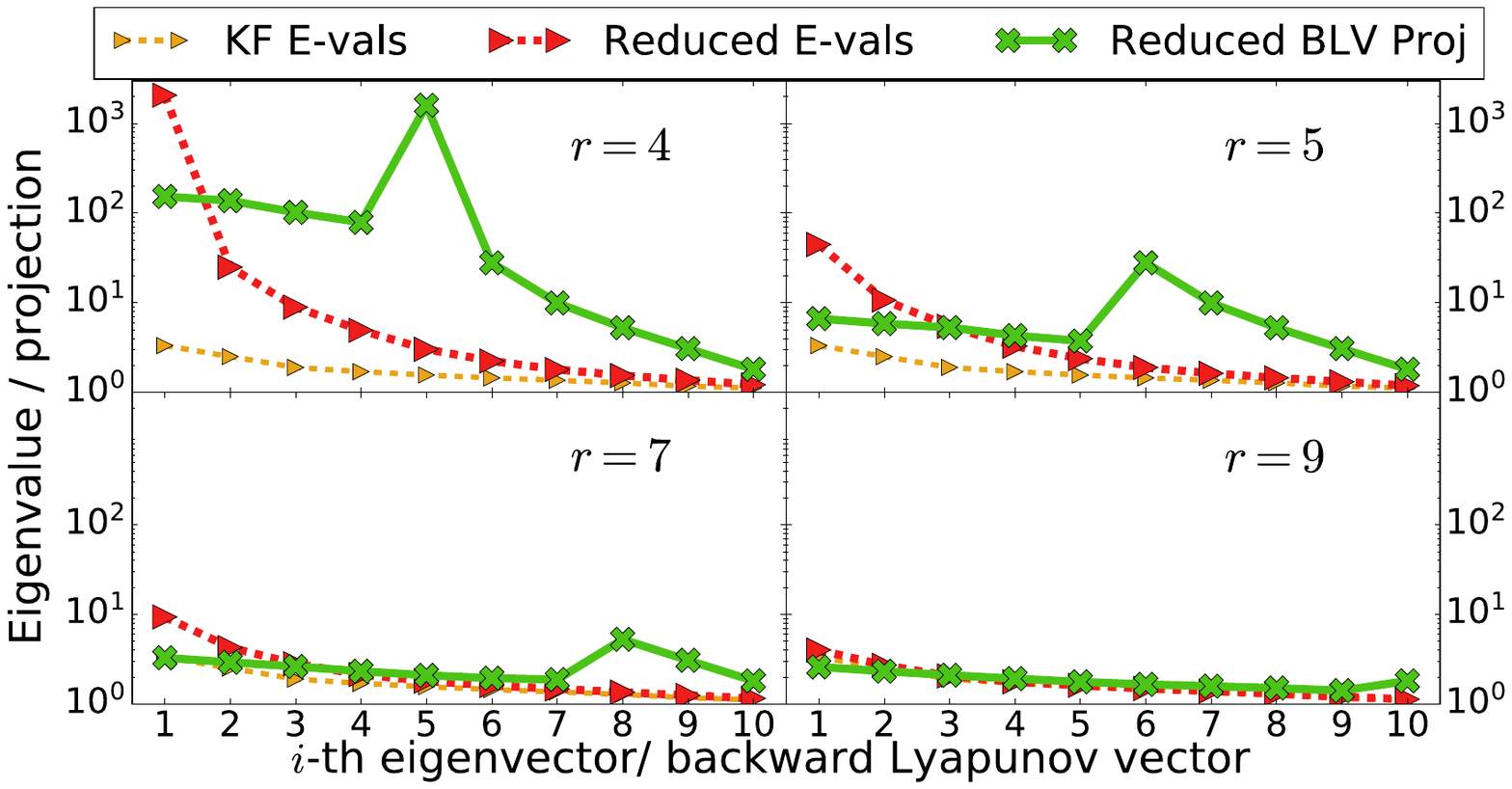}
\caption{Eigenvalues of the KF and the reduced-rank estimator covariances plotted with triangles.  Projection coefficients of the reduced-rank estimator covariance plotted with X's.\label{fig:AUSE_eigs}}
\end{figure}

Figure~\ref{fig:AUSE_eigs} presents an example from \cite{grudzien2018chaotic}, using the same tangent linear model from the 10-dimensional Lorenz-96 system as in Fig.~\ref{fig:bnd_v_lle}, and fixing each of $\mathbf{H}_k = \mathbf{R}_k = \mathbf{Q}_k = \mathbf{I}$.  In each window, the eigenvalues of the forecast error covariance matrix of the optimal full-rank KF (yellow) and of the ``exact'', reduced-rank estimator (red) are averaged over $10^5$ analysis cycles and plotted with triangles. By exact it is meant here that the full covariance equation is evolved analytically, including the covariance within the unfiltered trailing BLVs and their cross covariances with the leading filtered modes.
The rank, $r$, of the reduced-rank estimator is varied to examine the differences between the forecast error covariances arising from the optimal KF and the reduced-rank one when only the unstable-neutral subspace is corrected by the gain ($r=n_0=4$), the first stable mode is corrected by the gain ($r=5$) and so on. As suggested by Fig.~\ref{fig:bnd_v_lle}, correcting the first stable mode reduces the leading eigenvalue of reduced-rank estimator's forecast error covariance by an order of magnitude versus the case when only the unstable-neutral subspace has been corrected ({\it cf} the red curves between the two top windows).  

In each window the projection coefficients of the reduced-rank estimator's forecast error covariance into the basis of BLVs are also plotted (green line). For the full-rank optimal KF, the projection coefficients closely follow the eigenvalues and this is not shown due to redundancy. By contrast, it is clear that for the reduced-rank estimator the leading eigenvector is typically close to the first BLV that is contained in the null space of the reduced-rank gain, {\it i.e.} the first among the unconstrained directions. Figure~\ref{fig:AUSE_eigs} demonstrates thus a fundamental difference between the perfect model and stochastically forced models configurations: in stochastic systems, using a reduced-rank $n_0 \leq r< N_x $ Kalman estimator, the leading order forecast errors can actually lie in the stable, unfiltered modes. Those modes must be taken thus into account in the DA procedure by, for instance, appropriately enlarging the ensemble size or otherwise augmenting the span of the ensemble-based gain.

The exact recursion for the reduced-rank gain (red lines in Fig.~\ref{fig:AUSE_eigs}) requires resolving the full error covariance dynamics and therefore cannot be used practically for DA. Nevertheless, it was used here to render apparent, the upwelling phenomena: a fundamental source of uncertainty that is not captured by the standard reduced-rank KF (and thus almost all EnKF) recursion. This mechanism is ubiquitous whenever one solves for a reduced rank estimator and is present whenever the forecast error evolution can be well approximated by the tangent-linear dynamics (see Sect.~\ref{sec:3.2.2}). Notably, it also provides one basic, mathematically grounded, justification for using covariance inflation \citep{grudzien2018chaotic}, a powerful common fix used in ensemble-based DA (that are commonly rank-deficient by construction) to mitigate for sampling, and sometimes model, error \citep[see {\it e.g.}][their Section 4.4.2 and references therein]{carrassi2018data}. 
Finally, the exact recursion also demonstrates the asymptotic characteristics of the forecast error covariance when using a reduced-rank gain, in the absence of sampling error.  It is extremely important to note that in the exact error dynamics for the reduced rank estimator, the leading order forecast errors lie in the directions that are asymptotically stable but unfiltered.  This also highlights the importance of localization \citep{sakov2011relation} and gain-hybridization \citep{penny2017mathematical} as effective means for preventing the growth of forecast errors that lie outside of the ensemble span in the EnKF.

    \subsection{Nonlinear dynamics: the effect of chaos on the ensemble Kalman filter}
\label{sec:3.2}
        \subsubsection{Perfect and deterministic dynamics}
\label{sec:3.2.1}

The performance and the functioning mechanisms of the EnKF in nonlinear systems are studied with the aid of numerical simulations performed using the \verb|qgs| code platform \citep{demaeyer2020qgs, qgs_repo}.

We consider first a spectral 2-layer channel quasi-geostrophic atmospheric model. 
The Fourier modes decomposition is truncated at wavenumber $2$ in both meridional and zonal directions on a beta-plane, leading to a set of $20$ ODEs for the time evolution of the first $10$ components of the atmospheric streamfunction $\bpsi$, and temperature $\btheta$ \citep{reinhold1982dynamics}. The model dimension is $N_x=20$ and its spectrum of LEs includes $3$ positive and one neutral exponents, so that $n_0=4$.

The model is integrated with a time step of approximately $15$ minutes and is spun-up for $3$ years to ensure the solution has reached the model attractor. Afterward we initialise the DA experiments with the following protocol: a ``true'' trajectory is computed for $4,6$ years, and synthetic observations are generated by first sampling this trajectory at regular analysis time $t_k$, each separated by a time interval $\Delta t$, $t_{k+1} = t_k + \Delta t$. The observations are then obtained by adding a zero mean Gaussian random error sampled from $\mathcal{N}({\bf 0},\bR)$, with $\bR$ being the (assumed to be known) observation error covariance matrix. It is assumed that we observe the spectral components directly and that the full system is observed, implying that the observation operator is the identity matrix, $\bH=\bI_{N_x}\in\mathbb{R}^{N_x\times N_x}$. Although the former hypothesis cannot be met in practice (instrumental devices do not observe the spectral modes) and the latter rarely holds in high-dimensional applications, they are done here for the sake of clarity and will facilitate the study of the dynamical behaviour we intend to discuss. Furthermore, observational error is supposed to be spatially (in spectral space) uncorrelated with an amplitude proportional to the corresponding model variable's standard deviation, $\sigma_{{\rm md}}^i$,  $i=1,\ldots,N_x$. These imply $\bR=\sigma^{\rm \%}{\rm diag}(\sigma^1_{{\rm md}},\ldots,\sigma^{N_x}_{{\rm md}})$


Data assimilation is performed  using the EnKF-N, hereafter simply referred to as EnKF \citep{bocquet2011ensemble}. This EnKF belongs to the family of deterministic filters but it furthermore possesses the appealing property that the ensemble covariance multiplicative inflation is computed automatically as part of the DA process. Inflation is one of the unavoidable feature making the EnKF methods suitable for high dimensional problems \citep{carrassi2018data}. It comes under two ways known as multiplicative and additive inflation, that are often used together. We shall use multiplicative inflation alone because, as opposed to the additive version, it does not change the rank and span of the ensemble error covariance but it only inflates the matrix entries amplitude. This allows us to study the effect (if any) on the ensemble subspace (reflected into the ensemble covariance rank and span) that comes from the dynamics, without artifacts from the DA procedure. 
At each analysis step, the forecast anomaly matrix is inflated as $\alpha\bX^{{\rm f}}\gets\bX^{{\rm f}}$, $\alpha\ge1$.

We will study the properties of the EnKF ensemble subspace, its dimension and alignment to the unstable subspace of the underlying dynamics, and will investigate how those will relate to the skill of the EnKF. Following~\cite{bocquet2017four}, at each analysis time $t_k$, the alignment between the ensemble and the unstable-neutral subspaces, $\U_k$, is computed as 
\be
\label{eq:cos}
\cos^2(\theta^i_k) = \sum_{p=1}^{n_0} \cos^2(\theta^{i,p}_k) = \sum_{p=1}^{n_0} \frac{\left\{(\bu^p_k)^{\rm T}\bv^i_k\right\}^2}{\left\| \bv^{i}_k \right\|^2} \, \quad 1\le i\le N,
\ee
where $\theta^i_k \in [0,\pi]$ is the angle between the anomaly $\bv^i_k$ and $\U_k$, $\bu^p_k$ is the $p$-th BLV, and $N$ is the number of ensemble members.

\begin{figure}
    \centering
    \includegraphics[width=0.44\textwidth]{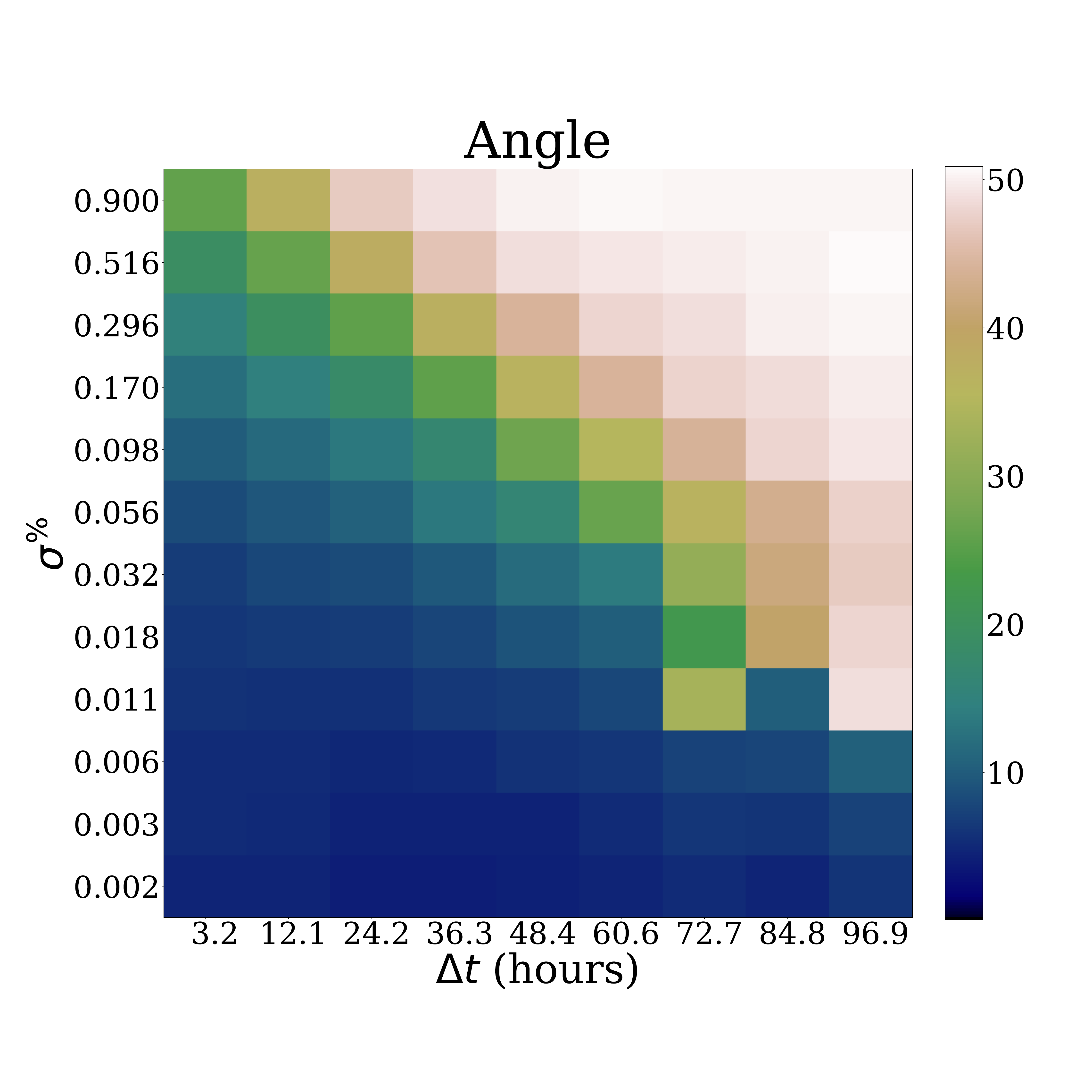}
    \includegraphics[width=0.44\textwidth]{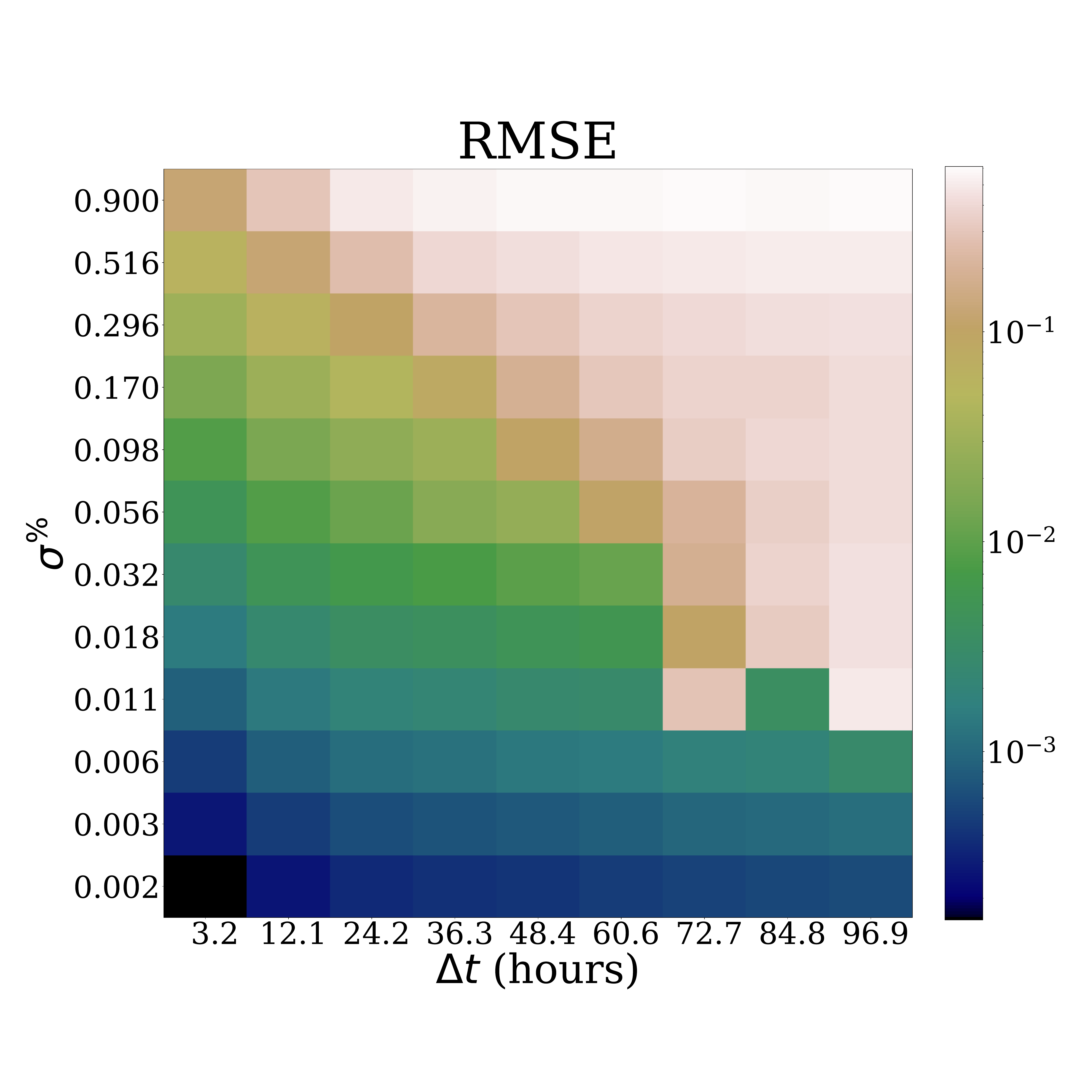}
    \caption{Atmospheric quasi-geostrophic model. Time- and ensemble-averaged angle, from Eq.~\eqref{eq:cos} (in degree), between the anomalies of the EnKF and the unstable–neutral subspace (left panel, shadow colours, in degree), and time averaged normalised RMSE of the EnKF analysis (right panel, shadow colours), both on the plane $(x,y) = (\Delta t,\sigma^{\rm \%})$. The set-up is $\bH = \bI_d$, $\bR=\sigma^{\rm \%}{\rm diag}(\sigma^1_{{\rm md}},\ldots,\sigma^{N_x}_{{\rm md}})$ and $N = 10$. Note that the logarithmic scale is used on the right panel.}
    \label{fig:qgs1}
\end{figure}

The RMSE of the EnKF analyses and the angle $\theta^i_k$ are shown on the plane $(x,y) = (\Delta t,\sigma^{\rm \%})$ in Fig.~\ref{fig:qgs1}. Values are averaged over $4$ years of simulated time after discarding the first $200$ analyses. The RMSE of the analyses is also averaged over all model variables, once the error on each variable is normalised with respect to the corresponding standard deviation $\sigma_{\rm md}$. The number of ensemble members is set to $N=10$. 

Figure~\ref{fig:qgs1} immediately shows the strong resemblance of patterns between the angle and the RMSE. In practice, whenever the observation time interval $\Delta t$ and error $\sigma^{\rm \%}$ are small enough then the angle between the two subspaces get smaller and the RMSE decreases. Similarly to what is discussed in \cite{bocquet2017four}, we also see here how increasing observation frequency is more effective in reducing the RMSE than reducing the observation error (see discussion in the Appendix of \cite{bocquet2017four}). This result also suggests that the use of a large number and frequent observations, like the ones produced by non-conventional systems ({\it e.g.} crowdsourcing) has the potential for improving analyses, despite that observational errors may be larger than in conventional measurement systems \citep{Nipen2020}.

\begin{figure}
    \centering
    \includegraphics[width=0.8\textwidth]{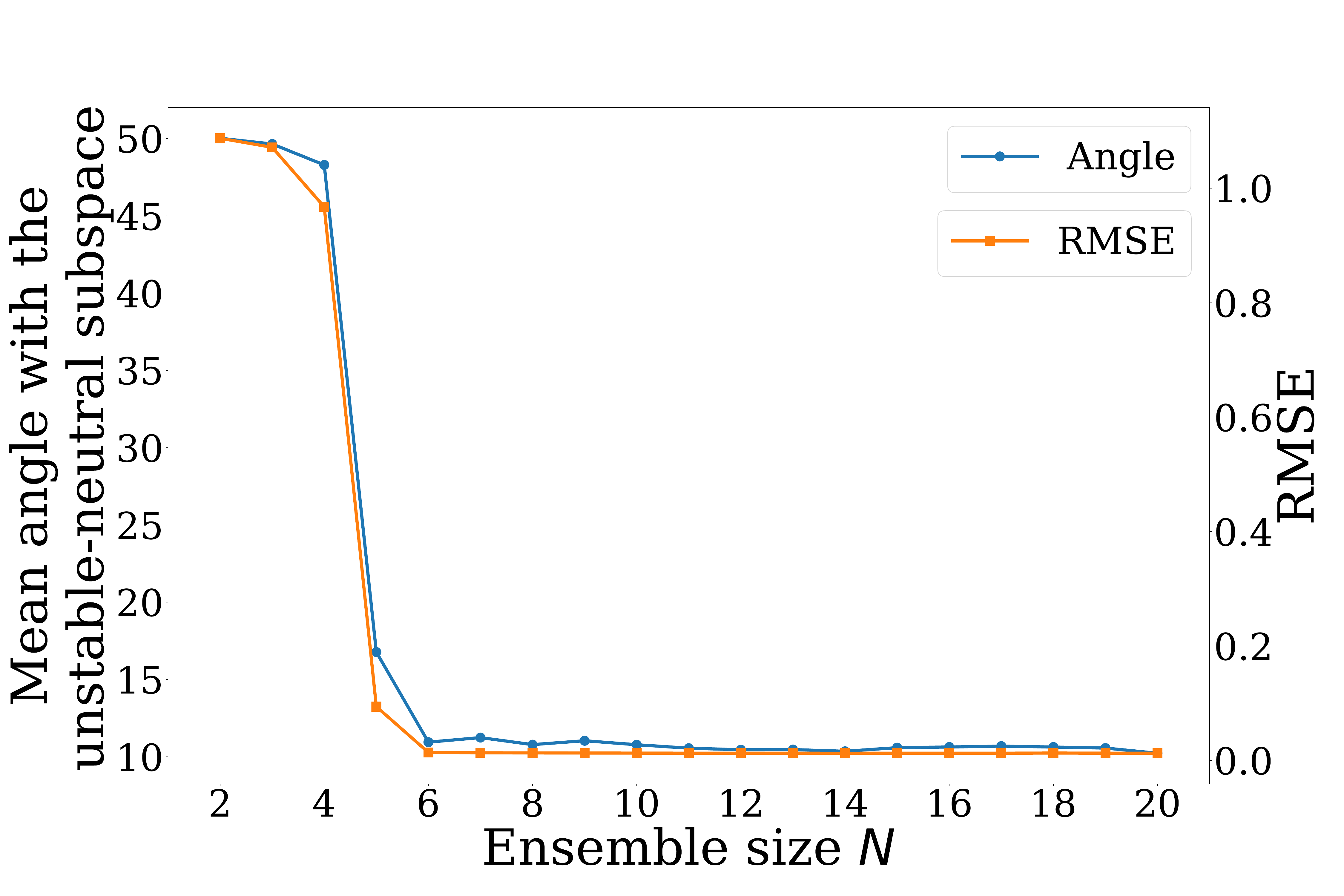}
    \caption{Atmospheric quasi-geostrophic model. Time- and ensemble-averaged angle, Eq.~\eqref{eq:cos} (in degree), between the ensemble anomalies of the EnKF and the unstable-neutral subspace as a function of the ensemble size $N$ (left $y$ axis), and the corresponding time-averaged RMSE of the EnKF (right $y$ axis). The set-up is $\bH = \bI_{N_x}$, $\bR = \bsigma\bsigma^{{\rm T}}$ with $\sigma^{\rm \%} = 0.08$ and $\Delta t= 12.11$ hours.}
    \label{fig:qgs2}
\end{figure}

A complementary picture of the relation between the filter skill and the subspaces alignment is provided in Fig.~\ref{fig:qgs2}, that displays the angle (left y-axis) and RMSE of the EnKF analysis (right y-axis) both against the ensemble size $N$. The remaining experimental set-up is $\bH = \bI_{N_x}$, $\bR = \bsigma\bsigma^{{\rm T}}$ with $\sigma^{\rm \%} = 0.08$ and $\Delta t= 12.11$ hours. Recalling that the unstable-neutral subspace has dimension $n_0=4$, Fig.~\ref{fig:qgs2} shows that as soon as $N\ge n_0+1$, {\it i.e.} the ensemble fully spans the unstable-neutral subspace, the RMSE suddenly reduces to very small values and it does not further decrease when $N$ is increased beyond $n_0+1$. The fact that the convergence occurs for $n_0+1$ instead of $n_0$ is due to the ensemble anomalies subspace to be at most of rank $N-1$ because one degree of freedom is removed when computing the ensemble mean. The behavior depicted in Fig.~\ref{fig:qgs2} is peculiar of the EnKF applied to chaotic dynamics and suggests a natural way to reduce computational cost by applying the EnKF with ``only'' $N=n_0+1$ members.   

\begin{figure}
    \centering
    \includegraphics[width=0.8\textwidth]{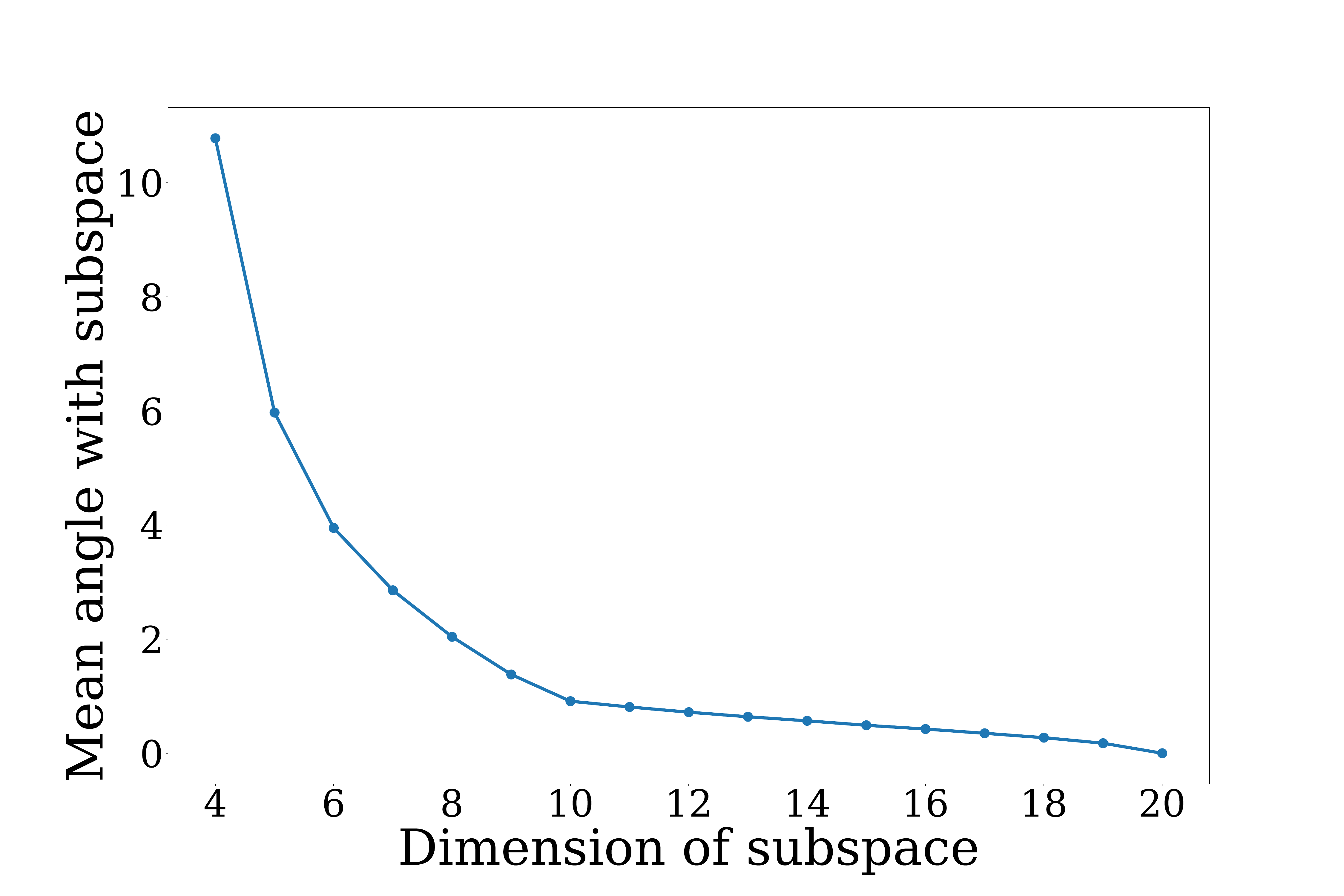}
    \caption{Atmospheric quasi-geostrophic model. Time- and ensemble-averaged angle (in degree) between the ensemble anomalies of the EnKF with $N=10$ members, and subspaces spanned by the BLVs of increasing dimensions. These subspaces are constructed by starting from the unstable-neutral subspace ($n_0=4$) and adding one by one stable directions ordered decreasingly by their Lyapunov exponent. The set-up is $\bH = \bI_d$, $\bR = \bsigma\bsigma^{{\rm T}}$ with $\sigma^{\rm \%} = 0.08$ and $\Delta t= 12.11$ hours.}
    \label{fig:qgs3}
\end{figure}

At the convergence, the angle between the ensemble and unstable-neutral subspace (of dimension $n_0=4$) is about $10$ degrees (see Fig.~\ref{fig:qgs2}): a remaining small portion of the ensemble subspace is projecting outside the unstable-neutral space. To investigate how large is such a portion, we compute the angle between an ensemble subspace with fixed $N=10$ and a Lyapunov/Oseledets subspace of increasing dimension beyond $n_0+1$; results are shown in Fig.~\ref{fig:qgs3}. The angle between the two subspaces decreases monotonically with the size of the subspace, eventually reaching zero once the full phase space ($n=20$) is considered. Interestingly, the rate of convergence is initially faster until approximately dimension $10$, and slower afterwards. This indicates that the additional projection beyond the asymptotic unstable-neutral subspace is largely confined to the less stable ({\it i.e.} close to neutral) directions, those that can often be locally unstable. This mechanism is a reminiscent of what was shown by \citep{grudzien2018chaotic} for stochastic systems and reviewed in Section~\ref{sec:3.1.2}. It suggests that the addition of few ensemble members beyond $n_0+1$ may lead to modest improvements in the filter skill, even though the long term performance is not much impacted.

To discuss the impact of multiple timescales on the ensemble Kalman filtering, we consider now the addition of a shallow-water ocean component to the previous atmospheric model. The atmospheric and oceanic models are coupled together through wind and radiative forcing as well as through heat exchanges, yielding the
Modular Arbitrary-Order Ocean-Atmosphere Model
\textsc{MAOOAM}~\citep{decruz2016maooam}. \textsc{MAOOAM} has the same $20$ ODEs of the atmospheric quasi-geostrophic model, and additional $16$ ODEs for the ocean. Amongst these $16$ equations, the first $8$ govern the time evolution of the first eights components of the oceanic streamfunction $\bpsi_{\rm o}$, while last $8$ are relative to the first eights components of the oceanic temperature anomaly $\delta\boldsymbol{T}_{\rm o}$. The model dimension is thus $N_x=36$ and its parameters are chosen such that a decadal low-frequency variability appears as a consequence of the coupling~\citep{vannitsemetal2015, decruz2016maooam}.

The model is integrated with a time step of approximately $15$ minutes and is spun-up for as many as $18,500$ years to ensure the solution has reached the model attractor, given the low-frequency ({\it i.e.} slow time-scale) of the model. From this spun-up trajectory we start DA experiments using the EnKF-N with the same protocol used for the atmospheric model detailed above. In this case the true trajectory lasts $185$ years, and it is again assumed that we observe the spectral components directly and completely. The EnKF is used in a strongly-coupled DA mode (see {\it e.g.} \cite{penny2017coupled}) so that, at analysis steps, atmospheric data can impact the ocean and vice-versa. MAOOAM has been already used as a prototypical coupled model to study coupled DA by \cite{penny2019strongly,tondeur2020temporal}. Again, observational error is assumed to be spatially (in spectral space) uncorrelated with an amplitude proportional to the corresponding model variable's standard deviation, $\sigma_{{\rm md}}^i$,  $i=1,\ldots,N_x$. These imply $\bR=\sigma^{\rm \%}{\rm diag}(\sigma^1_{{\rm md}},\ldots,\sigma^{N_x}_{{\rm md}}$).

\begin{figure}
    \centering
    \includegraphics[width=\textwidth]{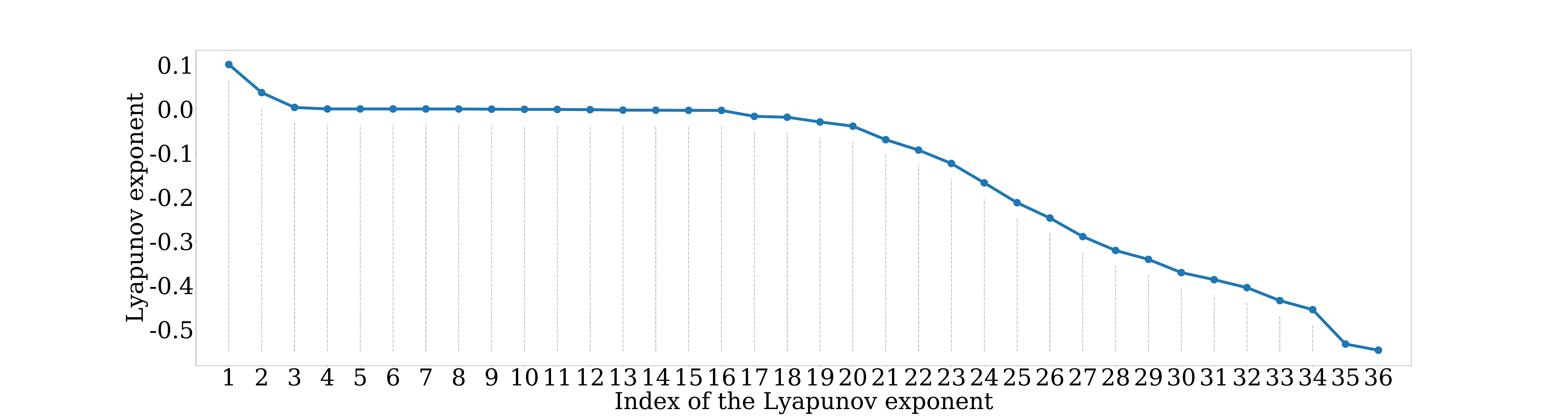}
    \caption{MAOOAM coupled Ocean-Atmosphere model. Time-averaged Lyapunov exponents in days$^{-1}$ computed along the true trajectory of the DA experiments.}
    \label{fig:maooam_spec}
\end{figure}

The spectrum of LEs of MAOOAM is displayed in Fig.~\ref{fig:maooam_spec} and presents some key features. 

With some arbitrariness, it can be decomposed into three subsets. A first subset of LEs corresponds to the unstable ($\lambda_i > 0$) and neutral ($\lambda_{n_0}=0$) directions. The dimension of the subspace spanned by these directions is $n_0=6$. A second subset of small negative LEs ($\lambda_i \in [-5 \times 10^{-3}, 0[ \; \mathrm{day}^{-1}$) corresponds to {\it nearly neutral directions}. The subspace they span has dimension $n_1=10$ and we hereafter define the {\it unstable-near--neutral subspace} as the one spanned by the first $n_0+n_1$ directions. The presence of these nearly-neutral directions amounts as a form of degeneracy of the neutral direction. Although the model is theoretically possessing only one single neutral mode ({\it cf} Sect.~\ref{sec:2}), it is computationally extremely difficult to distinguish it from other nearly neutral ones, and the model can be said to degenerate in the neutral direction in any practical sense.
We shall see how this feature has important consequences on the functioning and performance of DA. Finally, after a clear gap, the nearly neutral Lyapunov exponents are followed by the remaining stable directions which form the last subset of the spectrum. 

\begin{figure}
    \centering
    \includegraphics[width=0.3\textwidth]{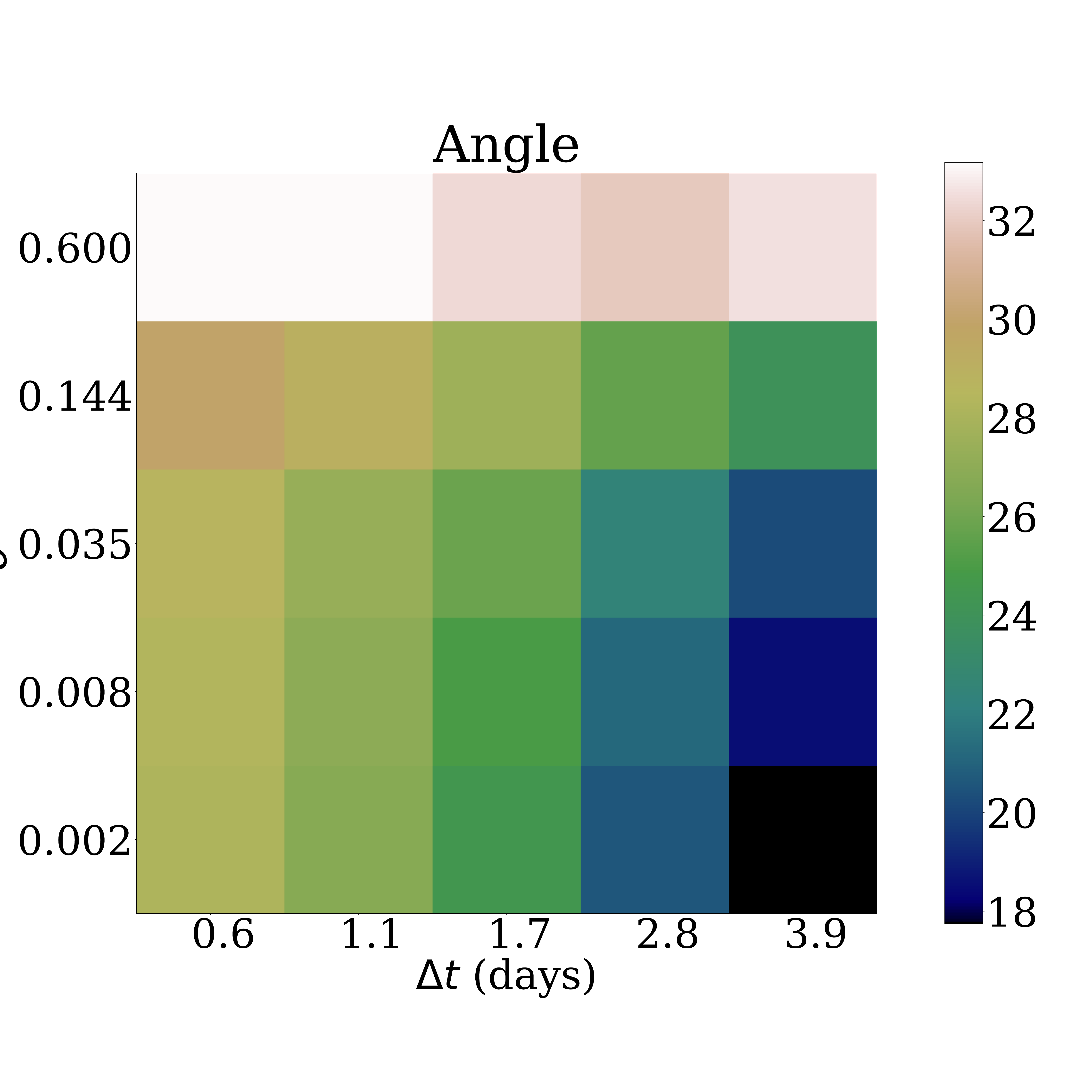}
    \includegraphics[width=0.3\textwidth]{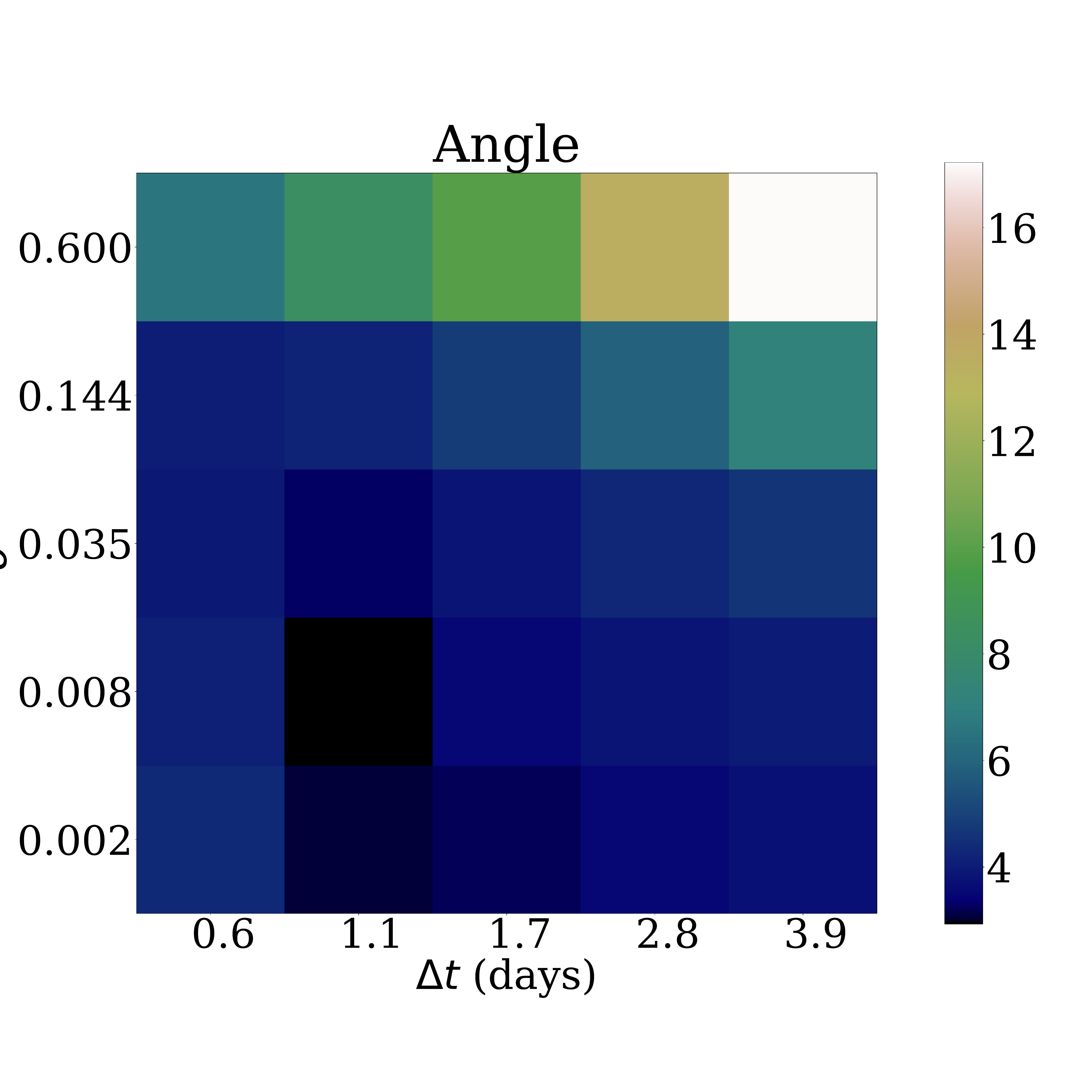}
    \includegraphics[width=0.3\textwidth]{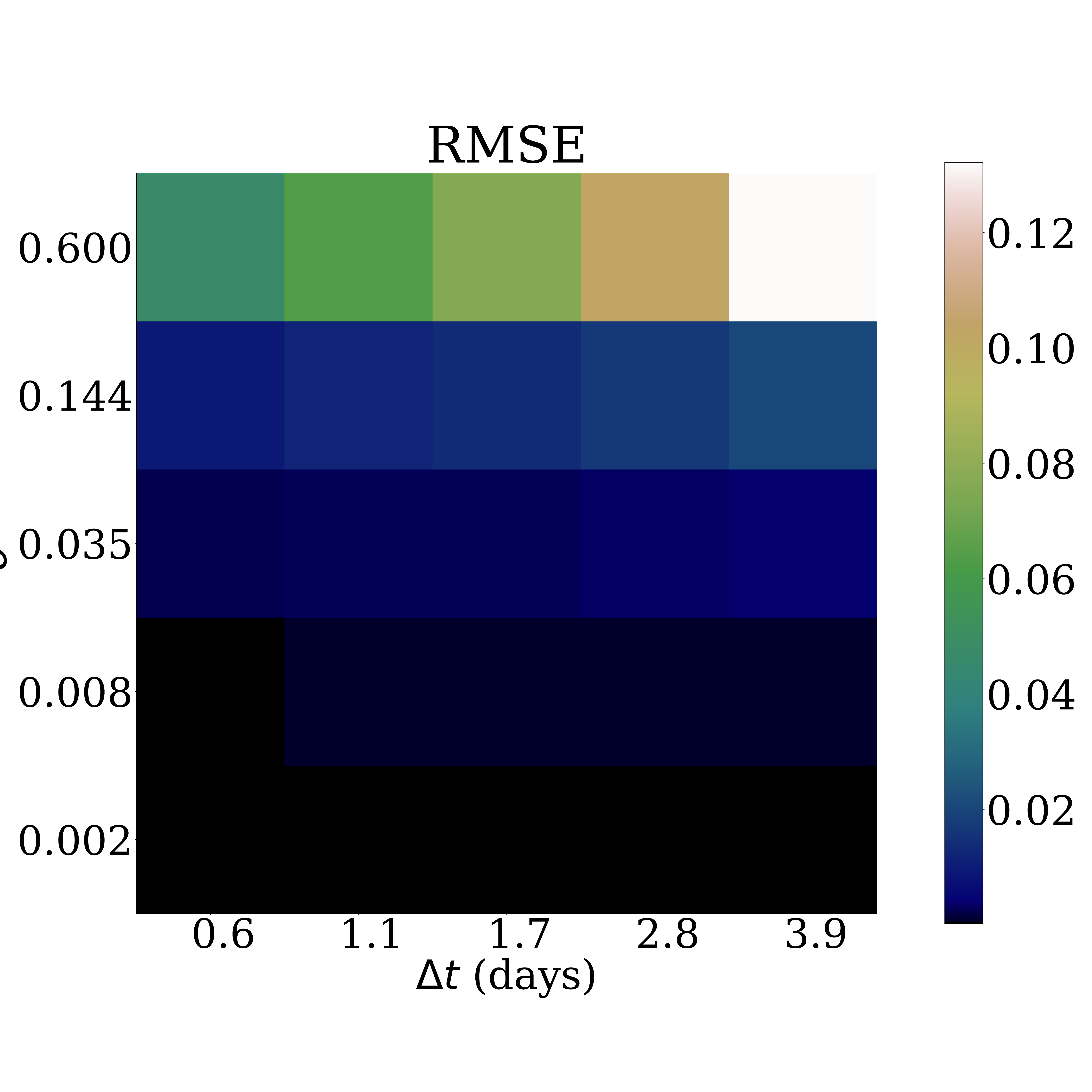}
    \caption{MAOOAM coupled Ocean-Atmosphere model. Time- and ensemble-averaged angle given by Eq.~\eqref{eq:cos} between the anomalies of the EnKF and the unstable–neutral subspace (left panel, shadow colours, in degree) or the unstable-near--neutral subspace (mid panel, shadow colours, in degree). The time averaged normalised RMSE of the EnKF analysis is also depicted  (right panel, shadow colours). All the figures are on the plane $(x,y) = (\Delta t,\sigma^{\rm \%})$. The set-up is $\bH = \bI_{N_x}$, $\bR=\sigma^{\rm \%}{\rm diag}(\sigma^1_{{\rm md}},\ldots,\sigma^{N_x}_{{\rm md}})$ and $N = 20$. The unstable-near--neutral subspace is defined as follows: it includes the subspace spanned by the unstable and neutral $n_0=6$ directions, but also an additional $n_1=10$ stable but near--neutral directions with LEs $\lambda_i \in [-5 \times 10^{-3}, 0[ \; \mathrm{day}^{-1}$.}
    \label{fig:maooam1}
\end{figure}

Similarly to Fig.~\ref{fig:qgs1}, we study the filter performance along with its ensemble subspace alignment to the unstable subspace in Fig.~\ref{fig:maooam1}. The figure shows the RMSE of the EnKF analyses and the angle $\theta^i_k$ given by Eq.~\eqref{eq:cos}, between the ensemble subspace and the unstable-neutral (left) and unstable-near--neutral (middle) subspace. Values are averaged over $185$ years of simulated time after discarding the first $100$ analyses. RMSE of the analysis is also averaged over all model variables, once the error on each variable is normalised with respect to the corresponding standard deviation $\sigma_{\rm md}$. The number of ensemble members is set to $N=20$. 

In contrast to what was observed in Fig.~\ref{fig:qgs1}, we see here that the pattern of the RMSE does not longer resemble the pattern of the angle between the ensemble and unstable-neutral subspace ({\it cf} left and right panels in Fig.~\ref{fig:maooam1}). Nevertheless, it bears great similarities with the pattern of the angle between the ensemble and unstable-near--neutral subspace ({\it cf} mid and right panels in Fig.~\ref{fig:maooam1}), suggesting that it is now this larger subspace (that includes the $n_1$ weakly stable modes) that contains most of the error. 

\begin{figure}
    \centering
    \includegraphics[width=0.8\textwidth]{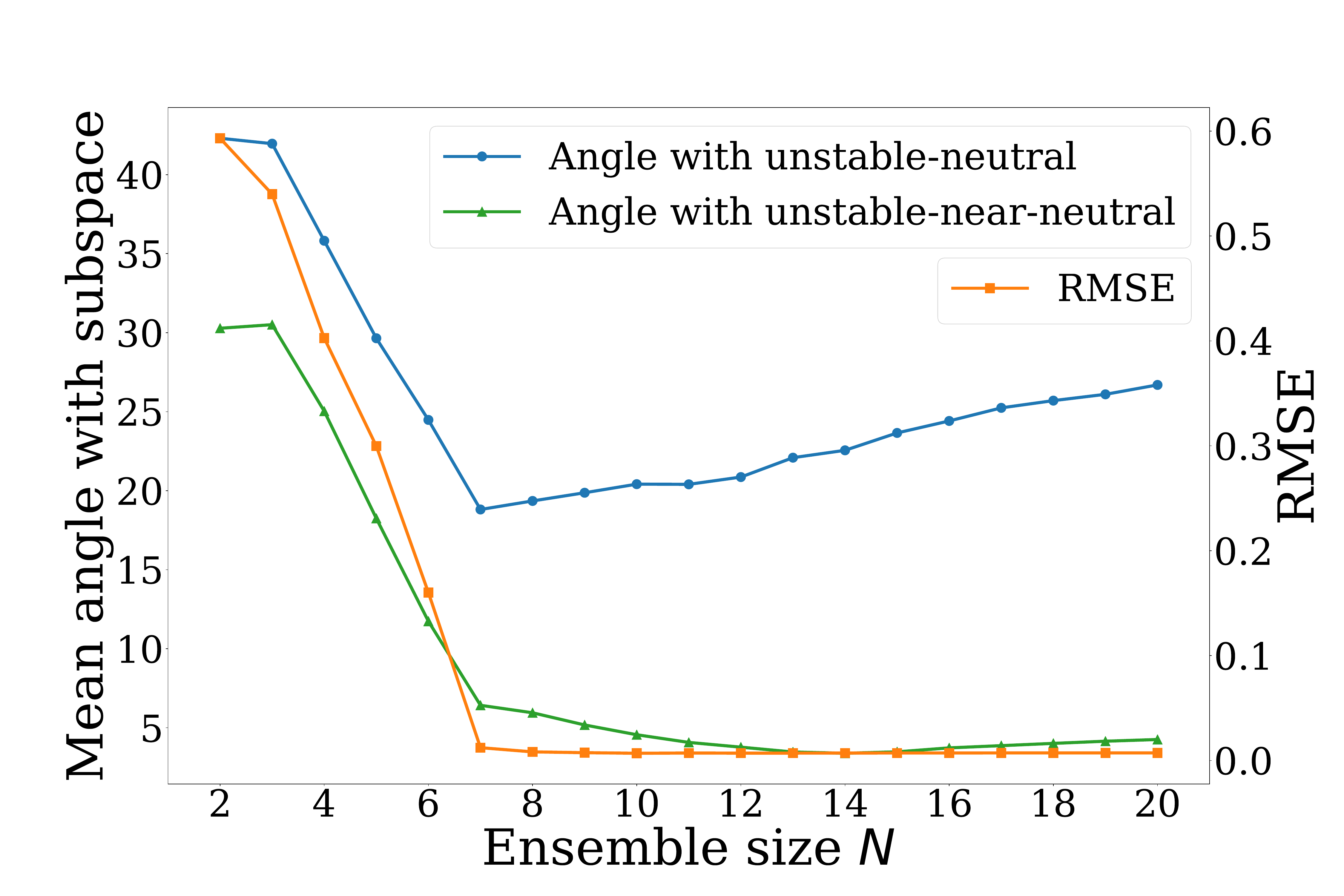}
    \caption{MAOOAM coupled Ocean-Atmosphere model. Time- and ensemble-averaged angle (in degree) between the ensemble anomalies of the EnKF and the unstable-neutral subspace or the unstable-near--neutral subspace as a function of the ensemble size $N$ (left $y$ axis), and the corresponding time-averaged RMSE of the EnKF (right $y$ axis). The set-up is $\bH = \bI_{N_x}$, $\bR=\sigma^{\rm \%}{\rm diag}(\sigma^1_{{\rm md}},\ldots,\sigma^{N_x}_{{\rm md}})$ with $\sigma^{\rm \%} = 0.08$ and $\Delta t= 1.68$ days.}
    \label{fig:maooam2}
\end{figure}

This is further confirmed by looking at Fig.~\ref{fig:maooam2} that, similarly to Fig.~\ref{fig:qgs2}, shows the angle (left y-axis) and RMSE of the analysis (right y-axis) against the ensemble size $N$. 
The set-up is $\bH = \bI_{N_x}$, $\bR=\sigma^{\rm \%}{\rm diag}(\sigma^1_{{\rm md}},\ldots,\sigma^{N_x}_{{\rm md}})$ with $\sigma^{\rm \%} = 0.08$ and $\Delta t= 1.68$ days.

\begin{figure}
    \centering
    \includegraphics[width=0.8\textwidth]{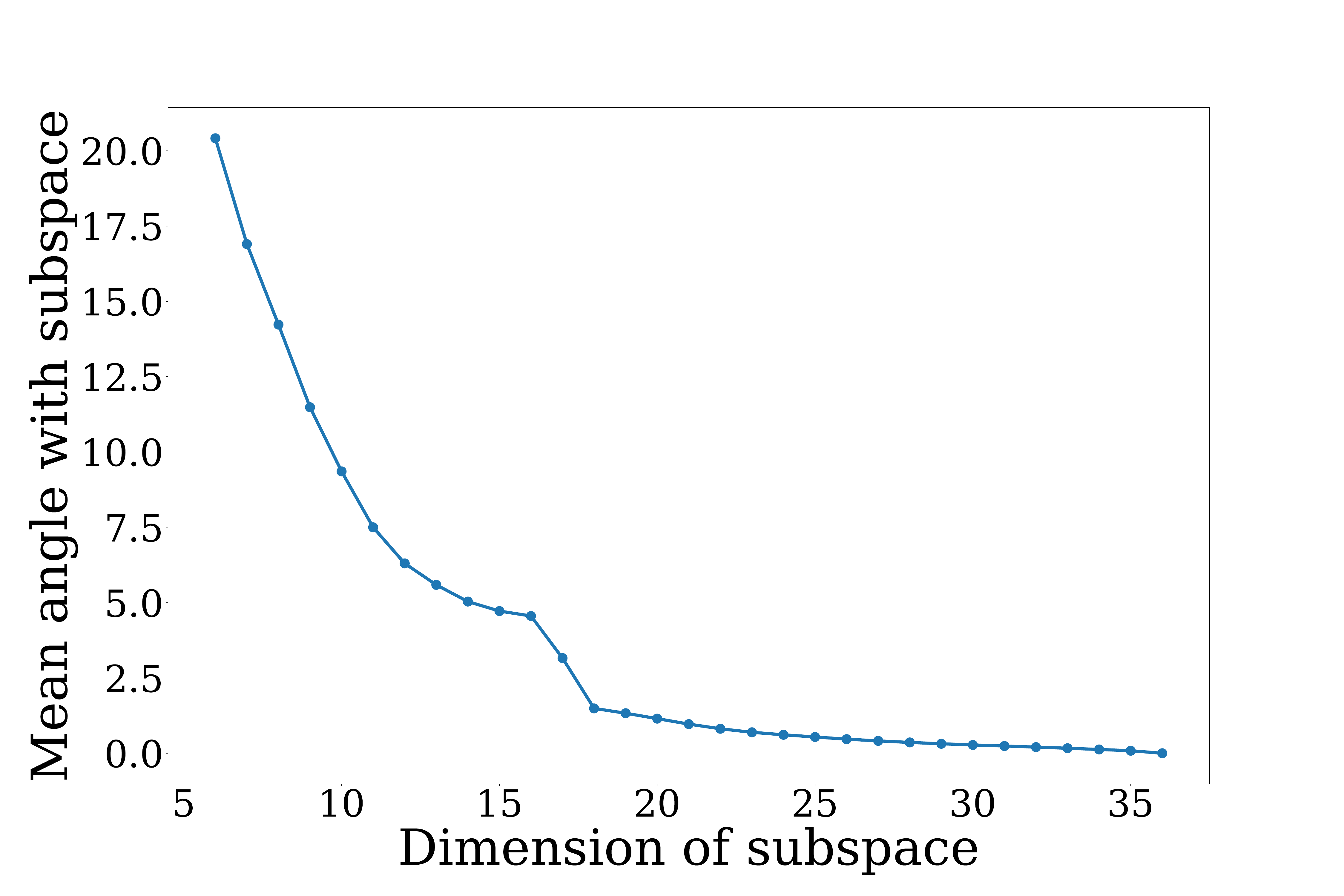}
    \caption{MAOOAM coupled Ocean-Atmosphere model. Time- and ensemble-averaged angle (in degree) between the ensemble anomalies of the EnKF with $N=10$ members, and subspaces spanned by the BLVs of increasing dimensions. These subspaces are constructed by starting from the unstable-neutral subspace ($n_0=6$) and adding one by one stable directions ordered decreasingly by their LE. The set-up is $\bH = \bI_{N_x}$, $\bR = \bsigma\bsigma^{{\rm T}}$ with $\sigma^{\rm \%} = 0.08$ and $\Delta t= 1.68$ days.}
    \label{fig:maooam3}
\end{figure}

The critical role of the $n_1$ weakly stable modes appears now evident when looking at Fig.~\ref{fig:maooam2} ({\it cf} green and blue lines). As opposed to the behaviour of the EnKF applied to the atmospheric model alone, or to the single scale Lorenz96 system used by~\cite{bocquet2017four}, we see that the ensemble subspace does not longer project much onto the unstable-neutral subspace (blue line with solid circles markers): when $N=n_0+1$ the angle reaches its minimum at around $20$ degrees (recall that it was about $10$ degrees for the atmospheric model, Fig.~\ref{fig:qgs2}), and it even further increases when $N>n_0+1$, indicating the presence of non-negligible projections outside the $n_0=6$ unstable-neutral directions. However, when the angle is computed with respect to the larger unstable-near--neutral subspace (green line with solid triangle markers), we retrieve the match with the RMSE curve (orange line with solid squares markers). A closer inspection further reveals that the angle decreases (the projection grows) fast until $N=n_0+1$: the first unstable-neutral $n_0=6$ modes still span most of the error. After this initial fast decrease, the angle keeps decreasing at a slower, yet monotonic, rate until approximately $N=15$ and stays almost constant afterwards. This result demonstrates undoubtedly the importance of the $n_1$ near--neutral modes. Although possibly asymptotically weakly stable, these directions span a small portion of the filter error that, if included in the ensemble subspace (by properly enlarging the ensemble size) leads to further improvement of the filter performance. This is finally emphasized in Fig.~\ref{fig:maooam3}, which shows the decrease of the ensemble-averaged angle between an ensemble of $N=10$ members and the subspaces of increasing dimension beyond the unstable-neutral one. In contrast to Fig.~\ref{fig:qgs3}, it indicates that the addition of $n_1$ extra ensemble members may lead to RMSE improvement that is far from negligible. Also, the gap observed around the value $n_0+n_1+1=17$ clearly shows that adding more members beyond the range of the near-neutral stability does not bring any benefit. This is a strong indication in favor of a cautious assessment of the ensemble size when working with coupled multi-scale dynamics and performing strongly-coupled DA. Furthermore, as elucidate by \cite{vannitsem2016} and \cite{tondeur2020temporal}, the near-neutral part of the spectrum in MAOOAM is directly connected to the coupling: including those directions within the ensemble subspace is paramount to propagate the data information content between ocean and atmosphere. This subspace also proved to be key in producing reliable ensemble forecasts in coupled ocean-atmosphere systems \cite{VannitsemDuan2020}.

        \subsubsection{Stochastic dynamics}
\label{sec:3.2.2}

Results in Sect.~\ref{sec:3.1.1} show that the optimal KF and the reduced-rank KF with gain restricted to the leading BLVs are asymptotically equivalent for linear, perfect-deterministic models.  Moreover, results in Sect.~\ref{sec:3.2.1} demonstrate that in weakly nonlinear error dynamics, with a perfect-deterministic model, these conclusions largely extend to the EnKF.  Yet the exact reduced-rank recursion introduced in Sect.~\ref{sec:3.1.2} evidenced important differences between the optimal and the reduced-rank formulations in the presence of model error for linear models. In this case, stochastic noise injected in asymptotically stable directions is not entirely damped out and remains finitely bounded. In addition, the upwelling process, albeit inherent to the reduced-rank formulation and not a consequence of model error, will move the constantly injected model noise from unfiltered to filtered directions. 

It is of interest thus to compare the differences between the full rank Kalman estimator, and both the standard and the exact reduced-rank KF recursions in a stochastically forced, nonlinear models. As a prototype, we use the model  defined by the nonlinear flow of the Lorenz-96 \citep{lorenz96} system with additive noise, and $N_x=40$. Suppose that $t_{k+1} = t_{k} + \Delta t$, so that the flow map taking all initial conditions to time $+\Delta t$ is defined $\phi_{\Delta t}(\x_k) = \x_{k+1}$.  If $\x^\mathrm{t}_k$ represents the true physical state at time $t_k$, we define the dynamical model analogously to Eq.~\eqref{eq:dynmodel} as a discrete, nonlinear map, 
$\x^\mathrm{t}_{k+1} = \phi_{\Delta t}\left(\x^\mathrm{t}_{k}\right) + \mathbf{w}_{k}.$

To avoid the interplay and superposition between sampling and model errors, and to be able to focus on the latter alone, instead of the EnKF we use here the EKF (see Sect.~\ref{sec:3} and \cite{jazwinski1970}).
The EKF estimates the forecast distribution for $\mathbf{x}^\mathrm{t}_k$ via the equation,
$\mathbf{x}^\mathrm{f}_{k+1} = \phi_{\Delta t}\left(\mathbf{x}^\mathrm{a}_k\right),$
taking the analysis mean at time $t_k$ to the forecast mean at time $t_{k+1}$, and by the linearized forecast equation for the covariance. In the following, the EKF propagates the full-rank covariance equation via the tangent-linear model defined along the mean equation and assimilates observations in all state components. On the other hand, both the standard and the exact reduced-rank formulations restrict the assimilation so that the image space of the gain is equal to the span of the leading $r$ BLVs defined along the tangent-linear model of the mean equation. The difference between the standard and the exact formulation is as follows: the standard formulation only estimates the error covariance in the span of the leading $r$ BLVs while the exact formulation simulates the entire covariance equation, estimating the free forecast covariance in the trailing modes and including its upwelling in the estimate of the uncertainty in the span of leading BLVs \citep[see proposition 1]{grudzien2018chaotic}

\begin{figure}[ht]
\center
\includegraphics[width=.49\linewidth]{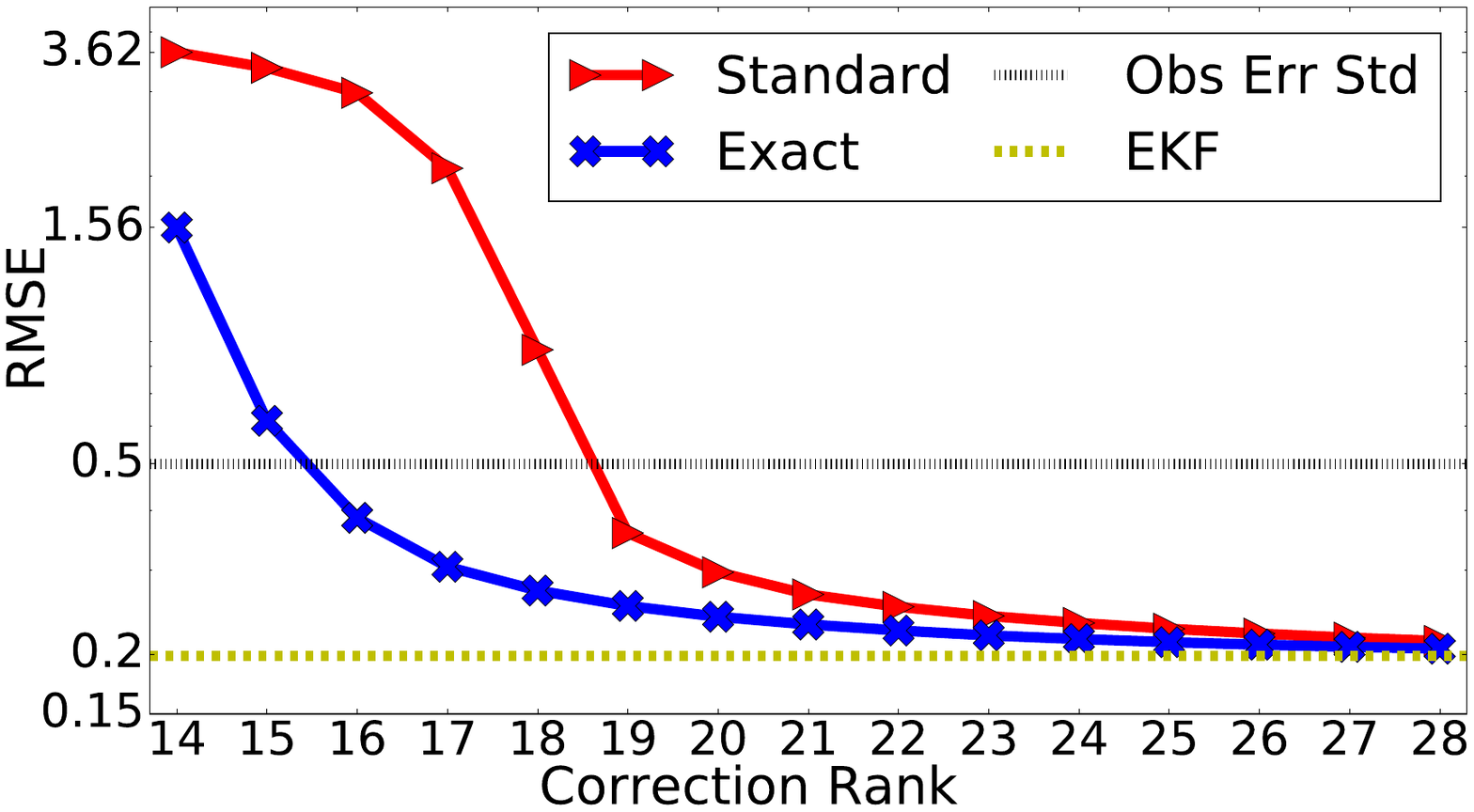}
\includegraphics[width=.49\linewidth]{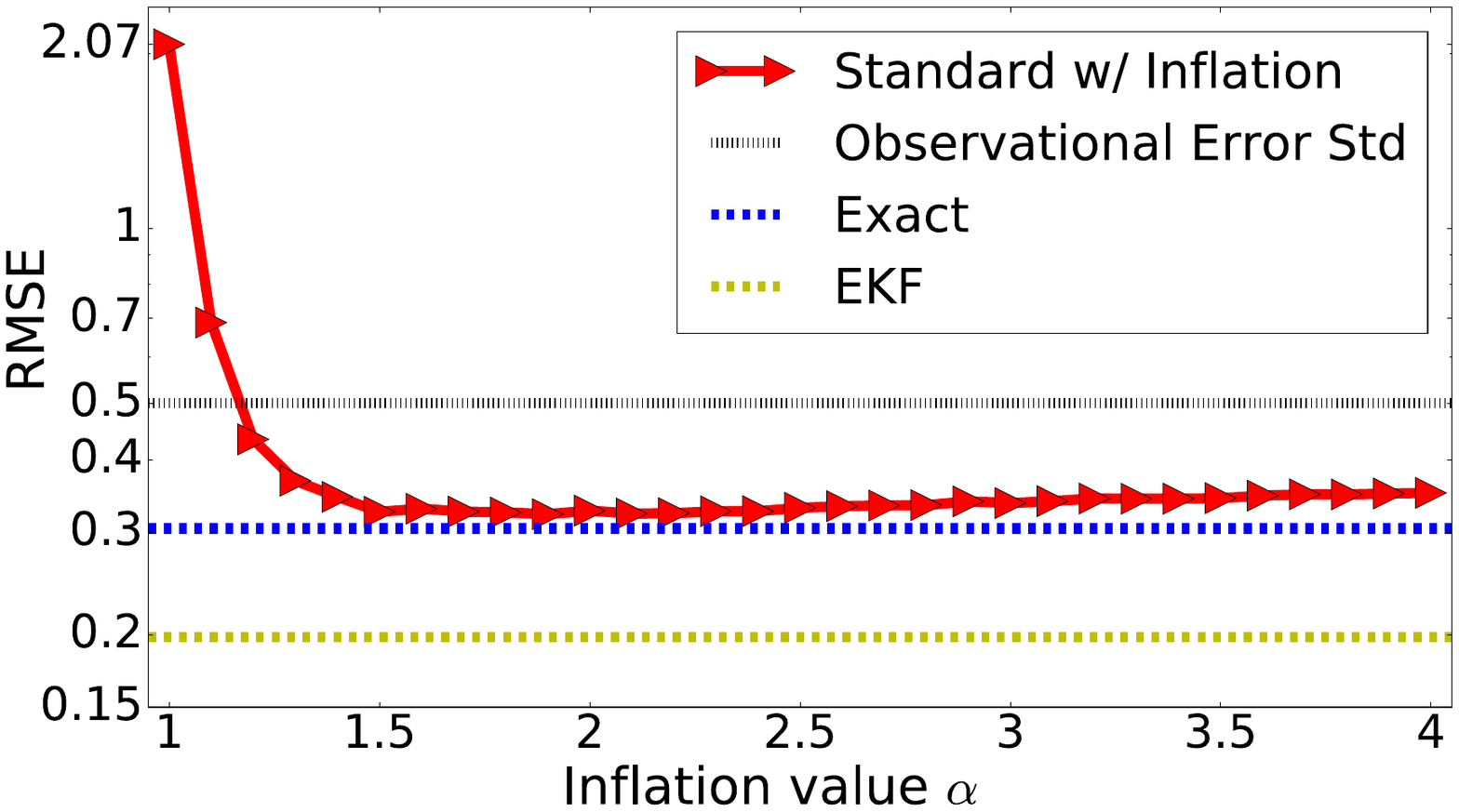}
\caption{\textbf{Left:} RMSE of the full-rank, standard reduced-rank and exact reduced-rank EKF.  The correction rank of the reduced-rank estimators is varied in the horizontal axis. \textbf{Right:} RMSE of the full-rank, standard reduced-rank and exact reduced-rank EKF.  Rank is fixed at $r=17$ and multiplicative inflation in the standard reduced-rank recursion is varied in the horizontal axis}
\label{fig:ekf_ause}
\end{figure}

Figure~\ref{fig:ekf_ause}, adapted from \cite{grudzien2018chaotic}, illustrates the differences in performance between the above described schemes. On the left, the average RMSE of the (i) \textit{full-rank}, the (ii) \textit{standard reduced-rank} and the \textit{exact reduced-rank EKF} formulations are plotted versus the rank of the reduced-rank gain over $10^5$ analyses. The model possesses $13$ positive LEs, therefore the reduced-rank EKFs under consideration have at least $r\ge n_0+1=14$.
The system is fully observed with $\bR=0.25\bI_{N_x}$.  As $r$ approaches $N_x=40$, the two reduced rank formulations converge to the full rank EKF, with performance considered optimal for a filter in this system.  However, for correction rank $r\leq 20$, there are substantial differences in performance between the standard formulation and that which includes the effect of the upwelling of error from the trailing BLVs: the exact formulation reaches both adequate and near-optimal performance with smaller $r$ than the standard recursion.  

On the right, Fig.~\ref{fig:ekf_ause} demonstrates the effect of multiplicative inflation on the standard recursion when the correction rank is fixed at $r=17$.  While multiplicative inflation greatly improves the performance of the standard recursion, this performance is actually bounded below by the RMSE of the exact formulation.  This example shows that, in addition to sampling error, multiplicative inflation can be used to remedy the inadequacies of the standard reduced-rank formalism which neglects the effects of dynamic upwelling.  This dynamic upwelling is a direct byproduct of the estimator being restricted to the span of the leading BLVs.  Such an estimator arises when, {\it e.g.}, the ensemble span aligns with the span of the leading BLVs in the EnKF, as demonstrated in Sect.~\ref{sec:3.2.1}. The efficacy of dynamic, multiplicative covariance inflation for treating the effect of model error separately from sampling error has also been demonstrated with statistical and optimization methods by, {\it e.g.}, \cite{mitchell2015accounting,raanes2015extending,raanes2019adaptive,sakov2018iterative,fillion2020iterative}. The dynamical upwelling derived in \cite{grudzien2018chaotic} provides an explanation of one of the mechanisms responsible for the need for covariance inflation.  

\section{Data assimilation for chaotic systems - How chaos became an opportunity}
\label{sec:4}

We present here two areas where the knowledge about the chaotic nature and properties of the model dynamics, have been used pro-actively to achieve a better track of the system of interest and to reduce the forecast error. There have been two ways how this has been accomplished. The former has to do with the design of methods to inform where and when additional fewer data would lead to a major improvement in the analysis and forecast skill. This is usually referred to as ``adaptive observation'' or ``target observation'', and is the content of Sect.~\ref{sec:4.1}. The second has to do with incorporating, within the DA process itself, the information about the unstable subspace, with the goal of achieving a computational economy while maximising the error reduction. This gave raise to a family of DA methods known as ``assimilation in the unstable subspace'', summarised in Sect.~\ref{sec:4.2}. 

Our treatments here are intentionally succinct and have the character of a review, but interested readers will find all the appropriate references to the original works.  
    \subsection{Targeting observations using the unstable subspace}
\label{sec:4.1}
Historically, one very important development in this type of dynamical analysis of the climate predictability / DA problem arose from the question of how to generate adaptive observation systems. By ``adaptive observations'' is meant here observations whose locations, time and type is chosen such that their impact on the state estimate or on the forecast skill is the largest.
It was well understood at the time that the growth of forecast errors was confined to a lower dimensional subspace of rapidly growing perturbations \citep{trevisan1995transient}. The Fronts and Atlantic Storm Track (FASTEX) program \citep{snyder1996summary}, a multinational collaboration to investigate the growth and development of frontal cyclones, in particular was motivated by these dynamical approaches to generate adaptive observation schemes that would target regions of rapid forecast error growth. Other similar international efforts have followed, some of them including actual field campaign of measurements: the THORPEX \citep{fourrie2006impact} and the Winter Storm Reconnaissance programs \citep{szunyogh2002propagation,hamill2013impact}.
Two main approaches were considered, the {\it forced singular vectors} \citep{buizza93,palmer1998singular}, and
the {\it bred vectors} \citep{toth1993ensemble,toth1997ensemble}, to identify the sensitivity areas of high forecast uncertainty.  

Forced singular vectors are generated by the right singular vectors of the forward-in-time, tangent-linear model resolvent $\mathbf{M}$. It can be seen from the earlier discussions in Sect.~\ref{sec:2} that the singular vectors can be interpreted as a finite-time approximation of the FLVs along a model forecast. They therefore indicate region where error will grow. On the other hand, the bred vectors are an ensemble-based approach to identify sensitivity regions.  Particularly, the breeding scheme simulates how the modes of fast growing error are maintained and propagated through the successive use of short range forecasts in weather prediction. The bred vectors are formed by initializing small perturbations of a control trajectory and forecasting these in parallel along the control. By successive rescaling of the perturbations amplitude back to a small value, this mimics the evolution of small perturbations under the tangent-linear model, and the span of these perturbations generically converges to the leading BLVs. Both of these approaches represent an early and intuitive way to utilizing the ergodic theory of chaotic dynamical systems in designing an effective adaptive observation scheme by targeting an unstable subspace in some form.

\cite{trevisan2004,uboldi2006} utilized the bred vectors / BLV analysis of \cite{toth1993ensemble,toth1997ensemble} and explicitly linked the methodology to Lyapunov stability theory. Importantly it was recognised that, instead of mimicking the error growth of the unforced (free) forecast, it was important to track the errors that develop within the DA cycle. This led to a modified version of the breeding approach known as Breeding on the Data Assimilation System (BDAS) \cite{Uboldi-2005,carrassi2007adaptive}. 
The BDAS scheme for adaptive observations is based on the principle of the support of the forecast error lying primarily in the span of the BLVs, with the bred vectors acting as a proxy for the explicit decomposition. In practice the locations of few adaptive observations were selected to be in the areas where the leading BDAS modes attained their local maxima. In experiments with an atmospheric quasi-geostrophic model, BDAS was used successfully to locate one additional observation at each analysis time of a 3DVar cycle, leading to a dramatic improvement of the analysis skill, compared to cases where either a fixed or a randomly located adaptive observation was assimilated \cite{carrassi2007adaptive}. 
    \subsection{Assimilation in the unstable subspace}
\label{sec:4.2}

Data assimilation has long been studied with the trade-off between accuracy and numerical efficiency as a goal. To this end a natural choice has been that of devising reduced-order schemes that focus the observational constraint on smaller, albeit crucial, part of the full dynamics. One of the most celebrated among those approaches is the assimilation in the unstable subspace (AUS) where the DA procedure is explicitly designed to track and control the unstable manifold of the dynamics, usually of much smaller dimension of the full phase space, thus aiming to a reduction in computational cost \citep{palatella2013a}. 

We have seen in Sect.~\ref{sec:2} that the full phase space of a chaotic dissipative dynamical system can be seen as split in a (usually much smaller) unstable–neutral subspace and a stable one \citep{Kuptsov2012}. For instance, \cite{carrassi2007adaptive} have shown how a quasi-geostrophic atmospheric model of $\mathcal{O}(10^5)$ degrees of freedom possesses an unstable–neutral subspace of dimension as small as $n_0= 24$.

We have furthermore seen that, in deterministic chaotic systems and under the linear regime of error evolution, the uncertainty in the state estimate converges to zero outside of the unstable-neutral subspace.
This phenomenon was at the core of the idea of AUS, whereby the unstable–neutral subspace (or a suitable numerical approximation of it) is explicitly used in the DA scheme to parametrise the description (both temporally and spatially) of the uncertainty in the state estimate \citep{trevisan2004assimilation,uboldi2006,carrassi2008a}. 

The AUS concept has been since then applied to different model scenarios and embedded within either KF-like or variational methods \citep{palatella2013a}. \cite{carrassi2008b} plugged AUS into a 3DVar cycle in such a way that the observations in the proximity of the leading BDAS mode's maximum were assimilated by imposing that the analysis increment follows the BDAS mode. This implied that the larger the estimated error growth with BDAS, the larger the analysis correction. The combined 3DVar-AUS was extraordinarily more accurate than the 3DVar when the same amount of data were assimilated. 
AUS was subsequently generalised and embedded into 4DVar (4DVar-AUS, \cite{trevisan2010}), and in an EKF, (EKF-AUS, \cite{trevisan2011kalman}). The forecast error covariance was projected so as to confine the analysis correction to the unstable-neutral subspace. Remarkably both reduced-rank formulations, 4DVar-AUS and EKF-AUS, showed superior skills than their full-rank counterparts. 

AUS relied on the assumption that errors evolve linearly. Going beyond this, \cite{palatella2015} presented an original way of mixing the contributions from the various Lyapunov vectors such that a quadratic expansions of the error is considered. This improved the performance of the standard EKF-AUS particularly in regimes of increasing nonlinearities. It remains however to be seen to which extent AUS concept could be used within fully nonlinear DA schemes. This is the subject of Sect.~\ref{sec:5.1} and of the references therein. Although AUS has been largely used in a perfect model scenario, \cite{PALATELLA201828} proposed a suitable modification that allows for incorporating parametric model errors. As proved in \cite{grudzien2018chaotic} and detailed in Sect.~\ref{sec:3.1.2} and \ref{sec:3.2.2} the use of AUS in chaotic systems forced by additive stochastic noise would necessarily require the additional inclusion of the asymptotically weakly stable modes.

A key caveat in all of the aforementioned applications of AUS is that one needs to compute in real time the BLVs to be used in the analysis. Therefore, while AUS proved capable to improve accuracy, it did not accomplishes a computational economy, unless the BLVs were all pre-computed and stored. Note however that the latter is not just a technological challenge given that the BLVs depends on the system's state and vice-versa if BLVs are to be used in the analysis update. Thus it is not straightforward to decouple their online estimation and use within the analysis. At the same time however, as we have seen in Sect.~\ref{sec:3}, AUS concept proved to be very powerful to understand, design and interpret the functioning of the KF and EnKF in chaotic systems.  

\section{Forward looking}
\label{sec:5}
    \subsection{AUS in non-Gaussian filter?}
\label{sec:5.1}
In this subsection we attempt to improve the performance of the (bootstrap) particle filter \citep[PF,][]{farchi2018survey} by AUS.
The underlying hypothesis is that observational components in the stable subspace
contribute little in the way of precision
(since nearby orbits within the stable subspace converge),
but a lot of noise.
Therefore, the investigation will explore whether
discarding observational information outside of the unstable subspace
can mitigate the acute collapse of weights
experienced by PFs in high-dimensional systems,
manifesting the ``{curse of dimensionality}''.
If so, this could be used to reduce the number of particles required,
which scales exponentially with some measure of the system size  \citep{snyder2008obstacles}.
A secondary objective is to investigate the effectiveness
of a few different methods of targeting observing systems
to the unstable subspace, potentially also reducing costs.

We perform synthetic DA experiments with the Lorenz-96 system.
The state size is set to $N_x=10$
and there is no dynamical noise ($\bQ=\bzero$).
Four different observation configurations targeting the unstable subspace are tested.
For each of them, observations are taken $0.2$ apart in time,
with independent noise of variance $1.5$.
Each experiment lasts for $10^5$ analysis cycles.
The RMSE averages of each method are tabulated
for a range of ensemble sizes and observation operator ranks, $N_y$,
and plotted as curves in Fig.~\ref{fig:bench_96}.
The plotted scores represent the lowest obtained among
a large number of tuning settings, selected for optimality at each point.
For the PF the tuning parameters are:
(i) the threshold for resampling,
which is triggered if the threshold is larger than
the effective ensemble size,
$\| \bw \|^{-2}$, where $\bw$ is the vector of weights,
and (ii) the bandwidth (scaling) of the regularizing post-resample jitter,
whose covariance is computed from the weighted ensemble.
For comparison the (symmetric square-root) EnKF \citep{hunt2004four} is also tested.
Its tuning parameters are
(i) the post-analysis inflation factor
and (ii) whether or not to apply random, covariance-preserving rotations \citep{sakov2008implications}.

\newcommand\bbC{\mathbf{\overline{C}}}
\newcommand\bbx{\mathbf{\overline{x}}}
\begin{figure}[ht]
\center
\includegraphics[width=.99\linewidth]{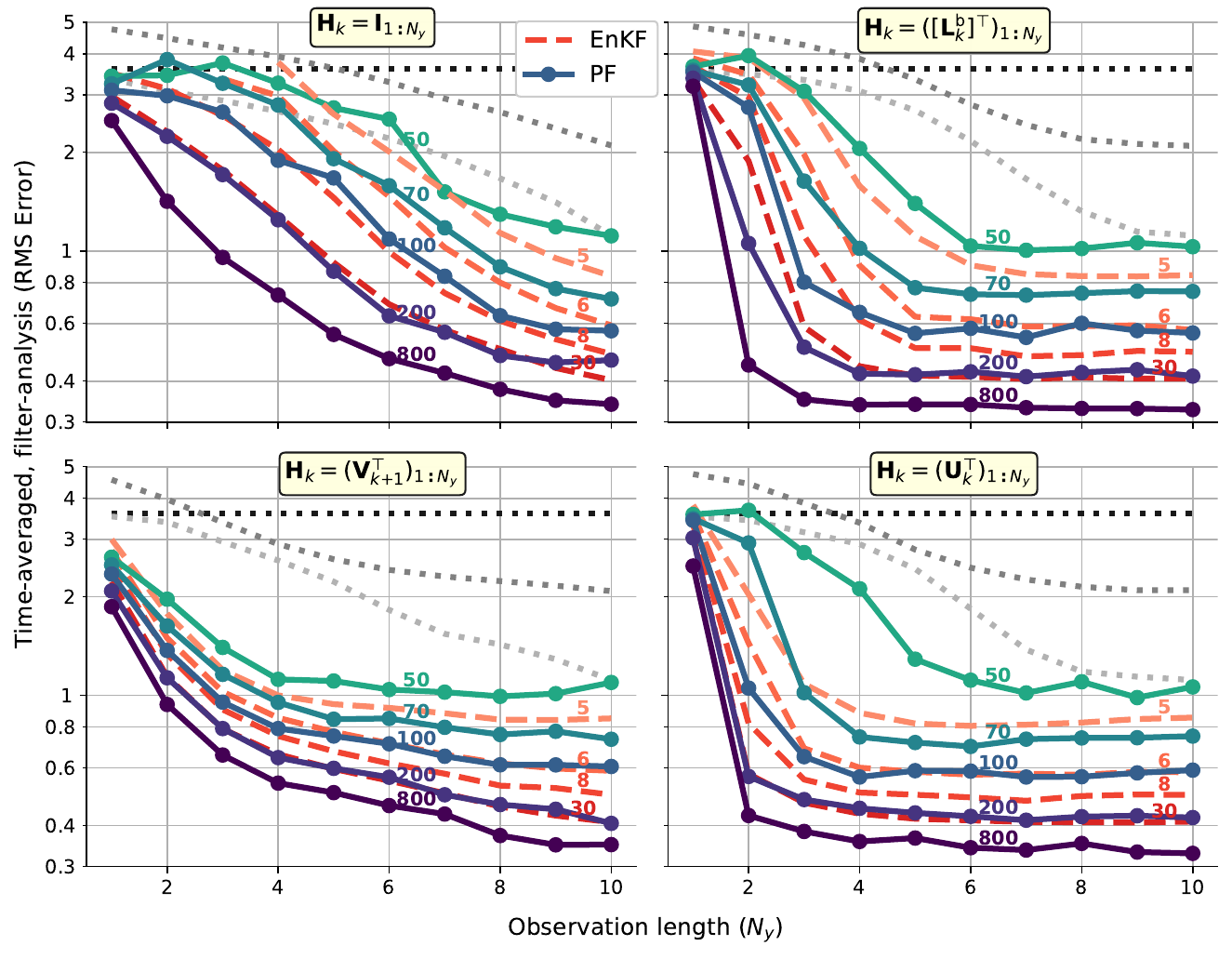}
\caption{
    Benchmarks of filter accuracy (RMSE)
    from synthetic DA experiments on the {Lorenz-96} system,
    plotted as functions of the observation dimension ($N_y$).
    Specifically, the observation operator
    consists of the $N_y$ leading rows of
    the transpose of
    the identity matrix (\emph{top-left}),
    the BLV matrix, $\bL_k^\textrm{b}$,
    computed by recursive QR decompositions (\emph{top-right}),
    the 1-cycle forward singular vectors, $\bV_{k+1}$ (\emph{bottom-left}),
    and 1-cycle backward singular vectors, $\bU_k$  (\emph{bottom-right}),
    all of which are defined via the fundamental matrix of
    the orbit of the (supposedly unknown) truth.
    The number of members/particles used for the ensemble Kalman filter (EnKF, \emph{dashed})
    and Particle filter (PF, \emph{solid}) is tagged above each line.
    No further improvement is obtained by increasing the
    ensemble size beyond
    $N=30$ for the EnKF
    and $N=800$ for the PF,
    for which
    the PF achieves better accuracy than the EnKF,
    as expected for nonlinear systems.
    The greyscale, dotted lines, included for context,
    show the performance of baseline methods,
    whose analysis estimates, $\xa$, are given by
    $
    \bbx$
    for Climatology (\emph{black}), 
    $
    \bbx + \bK(\bbC)\, [\by - \bbx]$
    for Optimal interp. (\emph{dark grey}), 
    $
    \xf + \bK(c \bI)\, [\by - \xf]$ 
    for 3D-Var (\emph{light grey}).
    Here, $\bbx$ and $\bbC$ are the mean and covariance of the
    (invariant measure of the) system dynamics,
    $\bK(\bC) = \bC \bH\T (\bH \bC \bH\T + \bR)^{-1}$
    is a gain matrix,
    $\xf$ is the forecast of the previous $\xa$,
    and $c$ is a scaling factor subject to tuning.
    %
    The plots show that 
    targeting observations to the directions of dynamical growth
    of the uncertainty is efficacious and that,
    for this task, $\bL_k^\textrm{b}$ and $\bU_k$ are similarly effective,
    and superior to $\bV_{k+1}$.
    Moreover, the RMSE performance of any method,
    also for the PF (which is our focus),
    never degrades with the inclusion of more observations.
}
\label{fig:bench_96}
\end{figure}

The top-left panel Fig.~\ref{fig:bench_96}
shows that the error decreases monotonically in $N_y$
when the observation operator are rows of the identity.
The top-right panel shows, by contrast,
that when observing the BLVs, $\bL_k^\textrm{b}$,
the PF with largest ensemble levels off at near-optimal performance with
as few as $N_y = n_0=4$ observations, only improving marginally thereafter.
The same trait can also be noted for lower $N$, albeit less clearly.
Also recall that a similar ``{levelling off}'' of RMSE around $n_0$ occurred
in Fig.~\ref{fig:qgs2} of Sect.~\ref{sec:3.2.1},
whose x-axis is $N$ (rather than $N_y$, as here).
This result demonstrates that
targeting observations to the directions of dynamical growth
of the uncertainty is efficacious.
Interestingly, the performance of any given DA method
is nearly independent of the observing system ({\it i.e.} panel) for $N_y = 10$.
This makes sense considering that all of
$\bI$, $\bL_k^\textrm{b}$, $\bV_{k+1}$, and $\bU_k$ are orthogonal,
{\it i.e.} equal up to a rotation.

In all panels of Fig.~\ref{fig:bench_96},
the RMSE score of the PF (as well as that of the EnKF),
for any ensemble size,
never degrades by the inclusion of more observations
(except for what we adjudge to be noise).
This feature was also found in experiments with
even smaller ensembles than shown,
and experiments with larger observation errors.
Thus it seems that the curse of dimensionality
is not mitigated by discarding observations.
Hence there is little to gain by eliminating
observation components outside of the unstable-neutral subspace,
apart for the potential computational efficiency in the case
that unstable-neutral has been pre-computed.

It should be noted that this finding runs counter to that
of \citet{maclean2019particle,beeson2020particle,potthast2019localized},
all of which report success in mitigating weight collapse
by reducing the observations to the leading components of $ \bL^\textrm{b}_k $ or some related matrix.
\citet{beeson2020particle} also tested observation operators
given by time-local backward and forward Lyapunov vectors,
defined through the singular value decomposition of the resolvent:
$\bM_k = \bU_k \bSigma_k \bV_{k}\T$.
They reported better targeting results with
$\bV_{k+1}$ than with $\bU_k$,
which in turn was more effective than $ \bL^\textrm{b}_k $.
Our results 
similarly show $\bU_k$ (bottom-right panel)
to be slightly more effective for targeting observations than
$\bL^\textrm{b}_k $ (top-right panel),
albeit only for intermediate ensemble sizes.
Both, however, are more effective than $\bV_{k+1}$ (bottom-left).
Moreover, also for $\bV_{k+1}$ and $\bU_k$,
we never observe lower RMSE scores when using fewer observation components.
There are some differences in the experimental setting;
notably
our system is deterministic,
and the jitter we apply is restricted to the ensemble subspace.
It is unclear if these differences
can account for the disparity in conclusions.

Why does PF performance not suffer from the inclusion of
further, visibly redundant, observations (contrary to our initial expectation)?
Consider the likelihood ({\it i.e.} weighting)
of particle $n \in [1,\ldots,N]$, at a given (implicit) time:
\newcommand*{\pdf}{\mathop{}\! p}
\providecommand{\abs}[1]{\lvert#1\rvert}
\providecommand{\norm}[1]{\lVert#1\rVert}
\newcommand{\bPi}{\mathbf{\Pi}}
\begin{align}
    \pdf(\by | \bx_n)
    &=
    \phi \bigl( \norm{\by - \bH \bx_n}^2_\bR \bigr) \, ,
    \label{eq:lklhdn}
\end{align}
where the norm is defined as $\norm{\by}_{\bR}^2 = \by\T\bR^{-1}\by$,
and $\phi$ is a radial density such that
Eq.~(\ref{eq:lklhdn}) represents an elliptical distribution,
for example a Gaussian.
Suppose for the sake of simplicity that
both the observation operator and its error covariance
are identity, {\it i.e.} $\bH=\bR=\bI$,
and let $\bPi$ be the matrix of orthogonal projection onto the
unstable-neutral subspace.
Now, assuming the PF controls the error in the unstable-neutral subspace,
it seems reasonable to assume, given the results for linear systems of Sect.~\ref{sec:3},
that the particles will converge onto the unstable-neutral subspace:
$(\bI - \bPi) (\bx_n - \bx) \to \bzero$,
at least as a first-order approximation.
Supposing this,
and by decomposing the norm into two orthogonal components,
it can be shown that
\begin{align}
    \norm{\by - \bH \bx_n}^2
    &=
    \norm{\bPi (\by - \bH \bx_n)}^2 + \norm{(\bI - \bPi) \bv}^2 \, ,
    \label{eq:norm_decomp}
\end{align}
where $\bv$ is the observation noise.
Thus, projecting the observations onto the unstable-neutral subspace,
which would eliminate
the last term
of Eq.~(\ref{eq:norm_decomp}),
merely reduces the data mismatch
by a \emph{constant} (with respect to $n$, the particle index).
Thus
\begin{align}
    \pdf(\bPi \by | \bx_n)
    &=
    c \pdf(\by | \bx_n) \, ,
    \label{eq:lklhd_proj}
\end{align}
for some $c>0$
which is rendered inconsequential by the subsequent weight normalization.
Hence, the inclusion of the remaining observation (outside the unstable-neutral subspace),
does not cause a higher variance in the weights,
nor the associated tendency to collapse/degeneracy.

The essence of the above reasoning is that the PF prior
has zero support, {\it i.e.} uncertainty, in the stable subspace,
and therefore ignores observational components,
\emph{including the noise}, in that subspace.
This should be contrasted with the situation for dynamical noise (model error),
which is ``{up-welled}'' from the stable to the unstable subspace,
as per Sect.~\ref{sec:3.1.2}.
This distinction exemplifies the difference between uncertainty addition
(by dynamical noise) and subtraction (by likelihood updates).

For nonlinear dynamical systems, as illustrated in Sect.~\ref{sec:3.2},
the particles do not neatly align with the unstable-neutral subspace.
Capturing this nonlinear aspect of the flow
is generally seen as an advantage of the PF.
It might be argued, though, that the observation noise is likely large compared to
the spread of the prior particles in the stable subspace,
and therefore the observations should be reduced by discarding the corresponding components.
However, as highlighted above,
the observational noise is constant in the particle index,
and therefore its amplitude does not constitute a particular source of weight variability.
Instead, the weight variability originates in the prior,
which was assumed to have low variability in this scenario.

\newcommand{\pinv}{\ensuremath{^+}}
\newcommand{\byh}{\mathbf{\hat{y}}}
\newcommand{\bQh}{\mathbf{\hat{Q}}}
\newcommand{\bQt}{\mathbf{\tilde{Q}}}
The analysis resulting in Eq.~(\ref{eq:lklhd_proj}) assumed that $\bH = \bR = \bI$ followed by $\bH = \bPi$.
In the case of the observation operator consisting of
the leading $N_y \ge n_0$ rows of the transposed BLV matrix,
$ \bH_k = (\mathbf{L}_k^\textrm{b})\T_{1:N_y}$,
the same conclusion can be derived along similar lines.
In the general case, for any $\bH$,
it is not obvious how to reduce the observations
so as to only measure the unstable-neutral subspace.
To accomplish this,
both \citet{maclean2019particle} and \citet{beeson2020particle}
apply the pseudo-inverse $ \bH\pinv $ before their reduction.
This can be costly if $ \bH $ is large.
A more practical approach is to reduce the observations
as $ \byh = (\bH \bQh)\pinv \by $,
or $ \byh = \bPi \by $ with $ \bPi = (\bH \bQh) (\bH \bQh)\pinv $,
with $\bQh = (\mathbf{L}_k^\textrm{b})\T_{1:N_y}$
In any case, re-doing the same derivation,
Eq.~(\ref{eq:lklhd_proj}) again follows,
including the same implications for the weights.

In summary, discarding observational information outside of the unstable subspace
does not yield improvements in the PF because it already
embodies this ``{flow-dependent}'' information.
A similar conclusion was also drawn for the iterative ensemble Kalman smoother
by \citet{bocquet2017four}.
Thus, while AUS is a powerful explanatory and diagnostics tool,
it is not obvious if it can be used to combat the curse of dimensionality for PFs.

Lastly, this section adds some clarification
to the influential paper by \citet{snyder2008obstacles},
whose conclusion
is sometimes taken to be
that the required ensemble size for PFs
scales exponentially with \emph{observation} size.
Our results rather indicate 
that the performance of a well-tuned PF will not deteriorate
with the inclusion of more observations (even if they are redundant).
In other words,
that the required ensemble size depends on
the rank of the \emph{state} space, or more precisely for chaotic dynamics,
the rank of the unstable-neutral subspace.
\citet{snyder2015performance} points out
that the ``{effective dimension}'' may be limited
by the observation size if this is smaller than the state size.
However, in case $N_y < n_0$, no filtering system using flow-dependent priors will be able to achieve
satisfactory performance, because the system is not sufficiently observed.
Of course, the question of observability is complicated by considering
time-dependent observation networks,
while localization can also be applied to alleviate the curse of dimensionality \citep{farchi2018survey}.

    \subsection{Data assimilation and random attractors }
\label{sec:5.2}

AUS and its theoretical extensions provide a framework to interpret the asymptotic inferences of the EnKF.  Provided that the forecast anomalies can be considered to be perturbations of the true physical state, and if their evolution can be approximated by the tangent-linear model along the true trajectory, the dynamics of the EnKF anomalies can be decomposed along the Oseledec spaces of the true physical state.  The stability and accuracy of the EnKF is largely determined by the ability of the ensemble-based gain to correct the growth of forecast errors in the unstable-neutral subspace, with respect to the uniform-complete observability of these modes. Additive noise complicates the description of the asymptotic forecast uncertainty as uncorrected forecast errors in the span of the stable BLVs may be bounded but impractically large.  The upwelling of such errors into the ensemble span furthermore necessitates covariance inflation to rectify the systematic underestimation of the forecast uncertainty in the standard KF-AUS recursion.  

Model stochasticity arises systematically in multiscale climate dynamics where there are large scale-separations between resolved and unresolved dynamic processes.  In the asymptotic limit of scale-separation, unresolved dynamics can be reduced to additive Gaussian noise due to the Central Limit Theorem; finite scale-separation in reality leads to non-Markovian memory terms in addition to additive stochastic forcing in the exact model reduction of a multiscale model, as in Mori-Zwansig formalism  \citep{gottwald2015stochastic}.  Several mathematically rigorous frameworks have been developed to model and simulate the effect of small-scale dynamics on the large-scale dynamics with stochastic parameterization, including averaging methods, perturbation methods and combinations of the two --- see, {\it e.g.}, the survey of approaches by \cite{demaeyer2018stochastic}.  

The theory of random dynamical systems offers a novel means of analysis of the DA cycle for multi-scale chaotic systems with large scale separations and model reduction error. Characterizing the asymptotic Bayesian posterior in terms of the properties of a random, nonlinear and ergodic attractor is a natural step forward in the philosopy of AUS.  Recent work suggests that the support of the posterior measure of the DA cycle can be asymptotically bounded by the support of the SRB measure in deterministic, nonlinear dynamical systems \citep{lea2018error}.  While this is an intuitively appealing result, the existence of an SRB measure in deterministic dynamics usually requires a hyperbolicity assumption which may not be appropriate in weather and climate, {\it e.g.} \citep{vannitsem2016}.  However, many of the theoretical challenges in showing the existence of SRB measures in deterministic dynamics are relaxed in a random dynamical systems setting.  Indeed, the Pesin entropy formula holds under very general assumptions for stochastic flows of diffeomorphisms \citep{liu2006smooth}[theorem 3.1 and discussion on page 127], establishing the link once again between the observed instability in the dynamics and the statistical properties of the invariant measure.

For such a dynamical interpretation of the DA cycle to be credible, the correct specification of random models in stochastic-physical systems is a primary concern; stochastically reduced models should be specified to preserve conservation laws and the original model's dynamics \citep{cotter2019numerically}. In addition to the correct stochastic model specification, important differences in the statistical properties of model forecasts of stochastic dynamical systems have been observed due to the discretization errors of certain low-order schemes. For example, \cite{frank2018note} develop an order 2.0 Taylor scheme to correct the bias in the drift term induced by the Euler-Maruyama scheme in their study system.  \cite{grudzien2020numerical} likewise find that the bias due to discritization error of the Euler-Maruyama scheme can be sufficient to cause filter divergence of the EnKF in the stochastically forced Lorenz-96 model.  \cite{grudzien2020numerical} emphasize the important role of efficient, weak numerical schemes for the simulation of ensemble-based forecasts.  Unlike strong convergent schemes, numerical schemes that converge in the weak sense can make reductions in the complexity of simulation by emphasizing the accuracy of the convergence of the ensemble to the forecast distribution rather than the accuracy of any individual ensemble member.

\section{Summary and Conclusion}
\label{sec:6}

Chaos is ubiquitous in natural, physical and laboratory systems. Scientists have long coped with this whenever attempting to model, predict or control such systems. Combining and confronting models with data is common in science and {\it data assimilation} (DA) is the term coined in the context of numerical weather prediction science to encompass the methods that perform such a combination. The outputs of DA is the improved representation of the system under study, and an estimate of the associated uncertainty. 

Inevitably, the sensitivity to initial conditions of chaotic systems, including state-dependence of the directions of error growth, is a challenge for DA. On the other hand, the energy dissipation typical of real systems implies a ``dimension reduction'' in that errors are confined within a subspace of the full system's phase space. 

The DA process requires furnishing a prior distribution, whose specification is a severe difficulty in high dimension. Gaussian methods reduce the complexity to that of estimating the prior mean and covariance. Yet, with the exceptionally high-dimensions of geophysical problems, a proper estimate, and storage of the prior covariance matrices is still challenging. Thus, suitable reduced-rank formulations should be used to lower the computational load while maintaining a good description of the errors about the full system.

In chaotic dissipative systems this goal can be achieved by monitoring the unstable-neutral subspace of the model dynamics and performing DA only within that subspace. This is the idea at the basis of the class of methods known as {\it assimilation in the unstable subspace} \citep{palatella2013a}, that were developed mainly in the years between 2004 to 2015; we reviewed them in Sect.~\ref{sec:4}. Despite reducing the problem size to that of the unstable-neutral subspace (of dimension $n_0\ll N_x$), AUS methods proved aptly skillful, very close to those of their more costly full-rank competitors. 

However, the reduction of cost achieved at the analysis steps is offset by the additional cost of monitoring and tracking the unstable-neutral subspace. This requires computing the Lyapunov vectors (usually the backward Lyapunov vectors), which implies computing the tangent linear model (or alternatively evolving an ensemble of bred perturbations mimicking the evolution under the tangent linear model) and a repeated QR matrix decomposition. For specific purposes, one can opt for tracking only a few dominant unstable modes. This was the case for early adaptive (targeted) observation results \citep{carrassi2007adaptive} or when AUS was used to complement a classical DA method \citep{carrassi2008b}.  

In parallel to the early developments of AUS, several studies with ensemble DA methods in chaotic dissipative systems were suggesting that a number of their key properties were related to the model instabilities, including the rank and span of the ensemble-based forecast error covariance, as well as the skill of the analysis \citep{sakov2008implications,carrassi2009,ng2011role,bocquet2014}. These results indicate that the ensemble anomalies automatically align with the unstable-neutral subspace, thus resulting in the analysis to be confined to it. 

To put this mechanism on a more rigorous theoretical footing, a stream of works have studied the relation between the unstable-neutral subspace in linear systems using the Kalman filter (KF) and Kalman smoother (KS) \citep{gurumoorthy2017,bocquet2017degenerate,bocquet2017four}. These works, reviewed in Sect.~\ref{sec:3.1}, have provided analytic proofs that the span of the error covariance matrices of the KF and KS tends asymptotically to the unstable–neutral subspace, independent of the initial condition ({\it i.e.} no matter the number of the ensemble members, provided it exceeds the size of the unstable-neutral subspace). For stochastic systems with additive noise it was proved that asymptotically weakly stable modes, that one might discard in deterministic systems, must be included and analytic bounds for the error were provided \citep{grudzien2018asymptotic}. 

How do these results hold for nonlinear systems? In the case of chaotic deterministic systems, this was studied in \cite{bocquet2017four} and further in Sect.~\ref{sec:3.2.1} of this chapter. It was numerically showed that an ensemble comprising at least as many members as the size of the unstable-neutral subspace plus one ($N\ge n_0+1$) is needed to achieve satisfactorily skill with the ensemble Kalman filter (EnKF). Section~\ref{sec:3.2.1} also considered the case of a coupled multiscale system with a quasi-degenerate spectrum of Lyapunov exponents. This originates in the presence of many close-to-zero exponents that are related to the coupling mechanisms \citep{vannitsem2016}. It is shown that their full inclusion in the ensemble design is needed to reduce the EnKF analysis error to a satisfactorily low level. 

The case of nonlinear stochastic chaotic systems with additive noise was studied in \cite{grudzien2018chaotic} and reviewed in Sect.~\ref{sec:3.2.2}. The section explains the role of the weakly stable modes already identified in linear systems, but also discovered the {\it upwelling mechanism} for which uncertainty is upwelled from unfiltered (stable) modes to filtered (unstable) ones. The upwelling phenomenon is not exclusive of nonlinear systems and it is in fact present in linear systems too. It provides an additional rationale to the use of multiplicative inflation, otherwise known by numerically evidence to be needed for a proper functioning of reduced-rank filters even in perfect model scenarios \citep[see {\it e.g.}][]{raanes2019adaptive}. 

An outlook at how this research may evolve is given in Sect.~\ref{sec:5}. In particular in Sect.~\ref{sec:5.1} we provide original results on the use of the AUS approach, {\it i.e.} exploiting the unstable-neutral subspace, in particle filters (PFs), a fully non-Gaussian DA method. Results indicate that targeting observations within the unstable-neutral subspace is very effective. However, by analogy with what was proved for the EnKF in Sect.~\ref{sec:3.2.1}, adding observations along the stable modes does not deteriorate the analysis. In the particle filter too, the particles automatically align along the unstable-neutral subspace so that the contribution from observations in its complement stable subspace is negligible. Our results shed also new insight on the scaling of the particle numbers needed to reach convergence. It is shown to depend on the size of the unstable-neutral subspace rather than the observation vector size. In Sect.\ref{sec:5.2} we surveyed how the novel concept of random attractors could offer new ways to further exploit the idea behind AUS on stochastic multi-scale systems with large scale separation. Moreover, we also described how to amend numerical integration schemes when doing ensemble DA on such systems, as a trade-off between accuracy and computational cost. 

It is important to recall that all of the results with the EnKF and PF that we have presented are obtained without the use of {\it localization} (see {\it e.g.} \cite{carrassi2018data}, their chapter 4.4, and \cite{farchi2018survey} for localisation in the EnKF and PF, respectively). While we are well aware of the dramatic positive impact of localization on the filters' skill, we intentionally did not use it as it artificially changes the ensemble-covariance rank and span, thus making it impossible to link them exclusively to the model instabilities. 

The use of the time-dependent unstable-neutral subspace to represent uncertainty in dynamical systems is still very appealing and potentially prone to success in a wider area than explored so far.
For instance, in a recent work by \cite{bocquet2020online} model error arise from parameter mispecification and the EnKF is applied to estimate simultaneously the (chaotic) model state and $N_p$ parameters. The EnKF is used in the state-augmentation formulation and the standard persistence model is adopted for the parameter dynamics. It was shown that the bound for the necessary ensemble size becomes $N \ge n_0+N_p+1$: $N_p$ additional members are required to infer the $N_p$ parameters. While the linear one-to-one relation between the number of parameters and that of the additional members is a consequence of the choice of a persistence model and will change if a different parameters dynamics is in place, this result further highlights how much the design of the EnKF is, and can be tied to the properties of the dynamical model. 

Future developments along these lines will unavoidably have to tackle the obstacle of the computing cost of the unstable-neutral subspace. We speculate that recent progress in the area of machine learning \citep{goodfellow2016deep} may help. Neural network surrogate models of chaotic systems have shown capabilities to reproduce the spectrum of the asymptotic Lyapunov exponents fairly well \citep{pathak2017using,BRAJARD2020101171}. It is matter of future investigations to explore the possibilities of machine learning algorithms that learn about the time-dependent instabilities from offline long model simulations and then assist the model in the prediction mode by providing a proxy of the unstable-subspace at each analysis time.

\begin{acknowledgement}
AC has been funded by the UK Natural Environment Research Council award NCEO02004. CEREA is member of Institut Pierre–Simon Laplace (IPSL).
PNR has been partly funded by DIGIRES, a project sponsored by industry partners and the PETROMAKS2 programme of the Research Council of Norway.
\end{acknowledgement}


\begin{thebibliography}{93}
\providecommand{\natexlab}[1]{#1}
\providecommand{\url}[1]{\texttt{#1}}
\expandafter\ifx\csname urlstyle\endcsname\relax
  \providecommand{\doi}[1]{doi: #1}\else
  \providecommand{\doi}{doi: \begingroup \urlstyle{rm}\Url}\fi

\bibitem[Adrianova(1995)]{adrianova1995introduction}
L.~Y. Adrianova.
\newblock \emph{Introduction to linear systems of differential equations}.
\newblock American Mathematical Soc., 1995.

\bibitem[Asch et~al.(2016)Asch, Bocquet, and Nodet]{asch2016data}
M.~Asch, M.~Bocquet, and M.~Nodet.
\newblock \emph{Data Assimilation: Methods, Algorithms, and Applications}.
\newblock SIAM, 2016.

\bibitem[Barreira and Pesin(2002)]{barreira2002}
L.~Barreira and Y.~B. Pesin.
\newblock \emph{Lyapunov Exponents and Smooth Ergodic Theory}.
\newblock Student Mathematical Library. American Mathematical Society, 2002.
\newblock ISBN 9780821829219.

\bibitem[Beeson and Namachchivaya(2020)]{beeson2020particle}
R.~Beeson and N.~S. Namachchivaya.
\newblock Particle filtering for chaotic dynamical systems using future
  right-singular vectors.
\newblock \emph{Nonlinear Dynamics}, June 2020.
\newblock \doi{10.1007/s11071-020-05727-y}.

\bibitem[Benettin et~al.(1980)Benettin, Galgani, Giorgilli, and
  Strelcyn]{benettin1980}
G.~Benettin, L.~Galgani, A.~Giorgilli, and J.~Strelcyn.
\newblock Lyapunov characteristic exponents for smooth dynamical systems and
  for {H}amiltonian systems; a method for computing all of them. {P}art 1:
  {T}heory.
\newblock \emph{Meccanica}, 15\penalty0 (1):\penalty0 9--20, 1980.

\bibitem[Bocquet(2011)]{bocquet2011ensemble}
M.~Bocquet.
\newblock Ensemble {K}alman filtering without the intrinsic need for inflation.
\newblock \emph{Nonlinear Processes in Geophysics}, 18\penalty0 (5):\penalty0
  735--750, 2011.

\bibitem[Bocquet and Carrassi(2017)]{bocquet2017four}
M.~Bocquet and A.~Carrassi.
\newblock Four-dimensional ensemble variational data assimilation and the
  unstable subspace.
\newblock \emph{Tellus A}, 69\penalty0 (1):\penalty0 1304504, 2017.

\bibitem[Bocquet and Sakov(2014)]{bocquet2014}
M.~Bocquet and P.~Sakov.
\newblock An iterative ensemble {K}alman smoother.
\newblock \emph{Q. J. R. Meteorol. Soc.}, 140:\penalty0 1521--1535, 2014.

\bibitem[Bocquet et~al.(2017)Bocquet, Gurumoorthy, Apte, Carrassi, Grudzien,
  and Jones]{bocquet2017degenerate}
M.~Bocquet, K.~S. Gurumoorthy, A.~Apte, A.~Carrassi, C.~Grudzien, and C.~K.
  R.~T. Jones.
\newblock Degenerate {K}alman filter error covariances and their convergence
  onto the unstable subspace.
\newblock \emph{SIAM/ASA Journal on Uncertainty Quantification}, 5\penalty0
  (1):\penalty0 304--333, 2017.

\bibitem[Bocquet et~al.(2020)Bocquet, Farchi, and Malartic]{bocquet2020online}
M.~Bocquet, A.~Farchi, and Q.~Malartic.
\newblock Online learning of both state and dynamics using ensemble kalman
  filters.
\newblock \emph{Foundation of Data Science}, 0:\penalty0 00--00, 2020.
\newblock \doi{10.3934/fods.2020015}.
\newblock Accepted for publication.

\bibitem[Brajard et~al.(2020)Brajard, Carrassi, Bocquet, and
  Bertino]{BRAJARD2020101171}
J.~Brajard, A.~Carrassi, M.~Bocquet, and L.~Bertino.
\newblock Combining data assimilation and machine learning to emulate a
  dynamical model from sparse and noisy observations: A case study with the
  lorenz 96 model.
\newblock \emph{Journal of Computational Science}, 44:\penalty0 101171, 2020.

\bibitem[Buizza et~al.(1993)Buizza, Tribbia, Molteni, and Palmer]{buizza93}
R.~Buizza, J.~Tribbia, F.~Molteni, and T.~N. Palmer.
\newblock Computation of optimal unstable structures for a numerical weather
  prediction model.
\newblock \emph{Tellus A}, 45\penalty0 (5):\penalty0 388--407, 1993.

\bibitem[Carrassi et~al.(2007)Carrassi, Trevisan, and
  Uboldi]{carrassi2007adaptive}
A.~Carrassi, A.~Trevisan, and F.~Uboldi.
\newblock Adaptive observations and assimilation in the unstable subspace by
  breeding on the data-assimilation system.
\newblock \emph{Tellus A}, 59\penalty0 (1):\penalty0 101--113, 2007.

\bibitem[Carrassi et~al.(2008{\natexlab{a}})Carrassi, Ghil, Trevisan, and
  Uboldi]{carrassi2008a}
A.~Carrassi, M.~Ghil, A.~Trevisan, and F.~Uboldi.
\newblock Data assimilation as a nonlinear dynamical systems problem: Stability
  and convergence of the prediction-assimilation system.
\newblock \emph{Chaos}, 18:\penalty0 023112, 2008{\natexlab{a}}.

\bibitem[Carrassi et~al.(2008{\natexlab{b}})Carrassi, Trevisan, Descamps,
  Talagrand, and Uboldi]{carrassi2008b}
A.~Carrassi, A.~Trevisan, L.~Descamps, O.~Talagrand, and F.~Uboldi.
\newblock Controlling instabilities along a {3DVar} analysis cycle by
  assimilating in the unstable subspace: a comparison with the {EnKF}.
\newblock \emph{Nonlinear Processes in Geophysics}, 15:\penalty0 503--521,
  2008{\natexlab{b}}.

\bibitem[Carrassi et~al.(2009)Carrassi, Vannitsem, Zupanski, and
  Zupanski]{carrassi2009}
A.~Carrassi, S.~Vannitsem, D.~Zupanski, and M.~Zupanski.
\newblock The maximum likelihood ensemble filter performances in chaotic
  systems.
\newblock \emph{Tellus A}, 61:\penalty0 587--600, 2009.

\bibitem[Carrassi et~al.(2018)Carrassi, Bocquet, Bertino, and
  Evensen]{carrassi2018data}
A.~Carrassi, M.~Bocquet, L.~Bertino, and G.~Evensen.
\newblock Data assimilation in the geosciences-an overview on methods, issues
  and perspectives.
\newblock \emph{WIREs Clim Change}, e535., 2018.
\newblock \doi{10.1002/wcc.535}.

\bibitem[Cotter et~al.(2019)Cotter, Crisan, Holm, Pan, and
  Shevchenko]{cotter2019numerically}
C.~Cotter, D.~Crisan, D.~D. Holm, W.~Pan, and I.~Shevchenko.
\newblock Numerically modeling stochastic lie transport in fluid dynamics.
\newblock \emph{Multiscale Modeling \& Simulation}, 17\penalty0 (1):\penalty0
  192--232, 2019.

\bibitem[De~Cruz et~al.(2016)De~Cruz, Demaeyer, and
  Vannitsem]{decruz2016maooam}
L.~De~Cruz, J.~Demaeyer, and S.~Vannitsem.
\newblock The modular arbitrary-order ocean-atmosphere model:
  \textsc{maooam}~v1.0.
\newblock \emph{Geoscientific Model Development}, 9\penalty0 (8):\penalty0
  2793--2808, 2016.
\newblock \doi{10.5194/gmd-9-2793-2016}.

\bibitem[Demaeyer and {De Cruz}(2020)]{qgs_repo}
J.~Demaeyer and L.~{De Cruz}.
\newblock Climdyn/qgs: qgs version 0.2.0 release, July 2020.
\newblock URL \url{https://doi.org/10.5281/zenodo.3941877}.

\bibitem[Demaeyer and Vannitsem(2018)]{demaeyer2018stochastic}
J.~Demaeyer and S.~Vannitsem.
\newblock Stochastic parameterization of subgrid-scale processes: A review of
  recent physically based approaches.
\newblock In \emph{Advances in Nonlinear Geosciences}, pages 55--85. Springer,
  2018.

\bibitem[Demaeyer et~al.(Submitted)Demaeyer, {De Cruz}, and
  Vannitsem]{demaeyer2020qgs}
J.~Demaeyer, L.~{De Cruz}, and S.~Vannitsem.
\newblock qgs: A flexible python framework of reduced-order multiscale climate
  models.
\newblock \emph{Journal of Open Source Software}, Submitted.
\newblock URL
  \url{https://raw.githubusercontent.com/openjournals/joss-papers/joss.02597/joss.02597/10.21105.joss.02597.pdf}.

\bibitem[Dieci and Van~Vleck(2002)]{dieci2002lyapunov}
L.~Dieci and E.~S. Van~Vleck.
\newblock Lyapunov spectral intervals: theory and computation.
\newblock \emph{SIAM Journal on Numerical Analysis}, 40\penalty0 (2):\penalty0
  516--542, 2002.

\bibitem[Dieci and Van~Vleck(2007)]{dieci2007lyapunov}
L.~Dieci and E.~S. Van~Vleck.
\newblock Lyapunov and sacker--sell spectral intervals.
\newblock \emph{Journal of dynamics and differential equations}, 19\penalty0
  (2):\penalty0 265--293, 2007.

\bibitem[Evensen(2009{\natexlab{a}})]{evensen2009}
G.~Evensen.
\newblock \emph{{D}ata {A}ssimilation: {T}he {E}nsemble {K}alman {F}ilter}.
\newblock Springer-Verlag Berlin Heildelberg, second edition,
  2009{\natexlab{a}}.

\bibitem[Evensen(2009{\natexlab{b}})]{evensen2009data}
G.~Evensen.
\newblock \emph{Data assimilation: the ensemble {K}alman filter}.
\newblock Springer Science \& Business Media, 2009{\natexlab{b}}.

\bibitem[Farchi and Bocquet(2018)]{farchi2018survey}
A.~Farchi and M.~Bocquet.
\newblock Review article: Comparison of local particle filters and new
  implementations.
\newblock \emph{Nonlinear Processes in Geophysics}, 25\penalty0 (4):\penalty0
  765--807, Nov. 2018.
\newblock \doi{10.5194/npg-25-765-2018}.

\bibitem[Fillion et~al.(2020)Fillion, Bocquet, Gratton, Gürol, and
  Sakov]{fillion2020iterative}
A.~Fillion, M.~Bocquet, S.~Gratton, S.~Gürol, and P.~Sakov.
\newblock An iterative ensemble kalman smoother in presence of additive model
  error.
\newblock \emph{SIAM/ASA Journal on Uncertainty Quantification}, 8\penalty0
  (1):\penalty0 198--228, 2020.
\newblock \doi{10.1137/19M1244147}.

\bibitem[Fourri{\'e} et~al.(2006)Fourri{\'e}, Marchal, Rabier, Chapnik, and
  Desroziers]{fourrie2006impact}
N.~Fourri{\'e}, D.~Marchal, F.~Rabier, B.~Chapnik, and G.~Desroziers.
\newblock Impact study of the 2003 north atlantic thorpex regional campaign.
\newblock \emph{Quarterly Journal of the Royal Meteorological Society},
  132\penalty0 (615):\penalty0 275--295, 2006.

\bibitem[Frank and Gottwald(2018)]{frank2018note}
J.~Frank and G.~A. Gottwald.
\newblock A note on statistical consistency of numerical integrators for
  multiscale dynamics.
\newblock \emph{Multiscale Modeling \& Simulation}, 16\penalty0 (2):\penalty0
  1017--1033, 2018.

\bibitem[Froyland et~al.(2013)Froyland, H{\"u}ls, Morriss, and
  Watson]{froyland2013computing}
G.~Froyland, T.~H{\"u}ls, G.~P. Morriss, and T.~M. Watson.
\newblock Computing covariant lyapunov vectors, oseledets vectors, and
  dichotomy projectors: A comparative numerical study.
\newblock \emph{Physica D: Nonlinear Phenomena}, 247\penalty0 (1):\penalty0
  18--39, 2013.

\bibitem[Ghil and Malanotte-Rizzoli(1991)]{ghil1991data}
M.~Ghil and P.~Malanotte-Rizzoli.
\newblock Data assimilation in meteorology and oceanography.
\newblock In \emph{Advances in geophysics}, volume~33, pages 141--266.
  Elsevier, 1991.

\bibitem[Goodfellow et~al.(2016)Goodfellow, Bengio, and
  Courville]{goodfellow2016deep}
I.~Goodfellow, Y.~Bengio, and A.~Courville.
\newblock \emph{Deep learning}.
\newblock MIT press, 2016.

\bibitem[Gottwald et~al.(2015)Gottwald, Crommelin, and
  Franzke]{gottwald2015stochastic}
G.~A. Gottwald, D.~Crommelin, and C.~Franzke.
\newblock Stochastic climate theory.
\newblock \emph{Nonlinear and Stochastic Climate Dynamics}, pages 209--240,
  2015.

\bibitem[Grudzien et~al.(2018{\natexlab{a}})Grudzien, Carrassi, and
  Bocquet]{grudzien2018asymptotic}
C.~Grudzien, A.~Carrassi, and M.~Bocquet.
\newblock Asymptotic forecast uncertainty and the unstable subspace in the
  presence of additive model error.
\newblock \emph{SIAM/ASA Journal on Uncertainty Quantification}, 6\penalty0
  (4):\penalty0 1335--1363, 2018{\natexlab{a}}.

\bibitem[Grudzien et~al.(2018{\natexlab{b}})Grudzien, Carrassi, and
  Bocquet]{grudzien2018chaotic}
C.~Grudzien, A.~Carrassi, and M.~Bocquet.
\newblock Chaotic dynamics and the role of covariance inflation for reduced
  rank {K}alman filters with model error.
\newblock \emph{Nonlinear Processes in Geophysics}, 25\penalty0 (3):\penalty0
  633--648, 2018{\natexlab{b}}.

\bibitem[Grudzien et~al.(2020)Grudzien, Bocquet, and
  Carrassi]{grudzien2020numerical}
C.~Grudzien, M.~Bocquet, and A.~Carrassi.
\newblock On the numerical integration of the lorenz-96 model, with scalar
  additive noise, for benchmark twin experiments.
\newblock \emph{Geoscientific Model Development}, 13\penalty0 (4):\penalty0
  1903--1924, 2020.

\bibitem[Gurumoorthy et~al.(2017)Gurumoorthy, Grudzien, Apte, Carrassi, and
  Jones]{gurumoorthy2017}
K.~S. Gurumoorthy, C.~Grudzien, A.~Apte, A.~Carrassi, and C.~K. R.~T. Jones.
\newblock Rank deficiency of {K}alman error covariance matrices in linear
  time-varying system with deterministic evolution.
\newblock \emph{SIAM Journal on Control and Optimization}, 55\penalty0
  (2):\penalty0 741--759, 2017.

\bibitem[Hamill et~al.(2013)Hamill, Yang, Cardinali, and
  Majumdar]{hamill2013impact}
T.~M. Hamill, F.~Yang, C.~Cardinali, and S.~J. Majumdar.
\newblock Impact of targeted winter storm reconnaissance dropwindsonde data on
  midlatitude numerical weather predictions.
\newblock \emph{Monthly weather review}, 141\penalty0 (6):\penalty0 2058--2065,
  2013.

\bibitem[Hunt et~al.(2004)Hunt, Kalnay, Kostelich, Ott, Patil, Sauer, Szunyogh,
  Yorke, and Zimin]{hunt2004four}
B.~R. Hunt, E.~Kalnay, E.~J. Kostelich, E.~Ott, D.~J. Patil, T.~Sauer,
  I.~Szunyogh, J.~A. Yorke, and A.~V. Zimin.
\newblock Four-dimensional ensemble {K}alman filtering.
\newblock \emph{Tellus A}, 56\penalty0 (4):\penalty0 273--277, 2004.

\bibitem[Jazwinski(1970)]{jazwinski1970}
A.~H. Jazwinski.
\newblock \emph{Stochastic Processes and Filtering Theory}.
\newblock Academic Press, New-York, 1970.

\bibitem[Kalman(1960)]{kalman1960}
R.~E. Kalman.
\newblock A new approach to linear filtering and prediction problems.
\newblock \emph{Journal of Fluids Engineering}, 82:\penalty0 35--45, 1960.

\bibitem[Kalnay(2003)]{kalnay2003atmospheric}
E.~Kalnay.
\newblock \emph{Atmospheric modeling, data assimilation and predictability}.
\newblock Cambridge {U}niversity {P}ress, 2003.

\bibitem[Kuptsov and Parlitz(2012)]{Kuptsov2012}
P.~V. Kuptsov and U.~Parlitz.
\newblock Theory and computation of covariant lyapunov vectors.
\newblock \emph{Journal of Nonlinear Science}, 22\penalty0 (5):\penalty0
  727--762, 2012.

\bibitem[Legras and Vautard(1996)]{legras1996}
B.~Legras and R.~Vautard.
\newblock A guide to lyapunov vectors.
\newblock In \emph{ECMWF Workshop on Predictability}, pages 135--146, Reading,
  United-Kingdom, 1996. ECMWF.

\bibitem[Liu and Qian(2006)]{liu2006smooth}
P.~D. Liu and M.~Qian.
\newblock \emph{Smooth ergodic theory of random dynamical systems}.
\newblock Springer, 2006.

\bibitem[Lorenz(1963)]{lorenz1963deterministic}
E.~N. Lorenz.
\newblock Deterministic nonperiodic flow.
\newblock \emph{Journal of the atmospheric sciences}, 20\penalty0 (2):\penalty0
  130--141, 1963.

\bibitem[Lorenz(1996)]{lorenz96}
E.~N. Lorenz.
\newblock Predictability: A problem partly solved.
\newblock In \emph{Proc. Seminar on predictability}, volume~1, 1996.

\bibitem[Maclean and Vleck(2019)]{maclean2019particle}
J.~Maclean and E.~S.~V. Vleck.
\newblock Particle filters for data assimilation based on reduced order data
  models, 2019.

\bibitem[Mitchell and Carrassi(2015)]{mitchell2015accounting}
L.~Mitchell and A.~Carrassi.
\newblock Accounting for model error due to unresolved scales within ensemble
  {K}alman filtering.
\newblock \emph{Quarterly Journal of the Royal Meteorological Society},
  141\penalty0 (689):\penalty0 1417--1428, 2015.

\bibitem[Ng et~al.(2011)Ng, McLaughlin, Entekhabi, and Ahanin]{ng2011role}
G.~H.~C. Ng, D.~McLaughlin, D.~Entekhabi, and A.~Ahanin.
\newblock The role of model dynamics in ensemble {K}alman filter performance
  for chaotic systems.
\newblock \emph{Tellus A}, 63\penalty0 (5):\penalty0 958--977, 2011.

\bibitem[Nipen et~al.(2019)Nipen, Seierstad, Lussana, Kristiansen, and
  Hov]{Nipen2020}
T.~N. Nipen, I.~A. Seierstad, C.~Lussana, J.~Kristiansen, and O.~Hov.
\newblock {Adopting Citizen Observations in Operational Weather Prediction}.
\newblock \emph{Bulletin of the American Meteorological Society}, 101\penalty0
  (1):\penalty0 E43--E57, 10 2019.
\newblock ISSN 0003-0007.
\newblock \doi{10.1175/BAMS-D-18-0237.1}.

\bibitem[Oljača et~al.(2018)Oljača, Bröcker, and Kuna]{lea2018error}
L.~Oljača, J.~Bröcker, and T.~Kuna.
\newblock Almost sure error bounds for data assimilation in dissipative systems
  with unbounded observation noise.
\newblock \emph{SIAM Journal on Applied Dynamical Systems}, 17\penalty0
  (4):\penalty0 2882--2914, 2018.
\newblock \doi{10.1137/17M1162305}.

\bibitem[Palatella and Grasso(2018)]{PALATELLA201828}
L.~Palatella and F.~Grasso.
\newblock The ekf-aus-nl algorithm implemented without the linear tangent model
  and in presence of parametric model error.
\newblock \emph{SoftwareX}, 7:\penalty0 28 -- 33, 2018.
\newblock ISSN 2352-7110.

\bibitem[Palatella and Trevisan(2015)]{palatella2015}
L.~Palatella and A.~Trevisan.
\newblock Interaction of {L}yapunov vectors in the formulation of the nonlinear
  extension of the {K}alman filter.
\newblock \emph{Physical Review E}, 91:\penalty0 042905, 2015.

\bibitem[Palatella et~al.(2013)Palatella, Carrassi, and
  Trevisan]{palatella2013a}
L.~Palatella, A.~Carrassi, and A.~Trevisan.
\newblock Lyapunov vectors and assimilation in the unstable subspace: theory
  and applications.
\newblock \emph{Journal of Physics A: Mathematical and Theoretical},
  46:\penalty0 254020, 2013.

\bibitem[Palmer et~al.(1998)Palmer, Gelaro, Barkmeijer, and
  Buizza]{palmer1998singular}
T.~Palmer, R.~Gelaro, J.~Barkmeijer, and R.~Buizza.
\newblock Singular vectors, metrics, and adaptive observations.
\newblock \emph{Journal of the Atmospheric Sciences}, 55\penalty0 (4):\penalty0
  633--653, 1998.

\bibitem[Pathak et~al.(2017)Pathak, Lu, Hunt, Girvan, and Ott]{pathak2017using}
J.~Pathak, Z.~Lu, B.~R. Hunt, M.~Girvan, and E.~Ott.
\newblock Using machine learning to replicate chaotic attractors and calculate
  lyapunov exponents from data.
\newblock \emph{Chaos: An Interdisciplinary Journal of Nonlinear Science},
  27\penalty0 (12):\penalty0 121102, 2017.

\bibitem[Penny et~al.(2019)Penny, Bach, Bhargava, Chang, Da, Sun, and
  Yoshida]{penny2019strongly}
S.~Penny, E.~Bach, K.~Bhargava, C.-C. Chang, C.~Da, L.~Sun, and T.~Yoshida.
\newblock Strongly coupled data assimilation in multiscale media: Experiments
  using a quasi-geostrophic coupled model.
\newblock \emph{Journal of Advances in Modeling Earth Systems}, 11\penalty0
  (6):\penalty0 1803--1829, 2019.

\bibitem[Penny(2017)]{penny2017mathematical}
S.~G. Penny.
\newblock Mathematical foundations of hybrid data assimilation from a
  synchronization perspective.
\newblock \emph{Chaos: An Interdisciplinary Journal of Nonlinear Science},
  27\penalty0 (12):\penalty0 126801, 2017.

\bibitem[Penny and Hamill(2017)]{penny2017coupled}
S.~G. Penny and T.~M. Hamill.
\newblock Coupled data assimilation for integrated earth system analysis and
  prediction.
\newblock \emph{Bulletin of the American Meteorological Society}, 98\penalty0
  (7):\penalty0 ES169--ES172, 2017.

\bibitem[Pikovsky and Politi(2016)]{pikovsky2016lyapunov}
A.~Pikovsky and A.~Politi.
\newblock \emph{Lyapunov exponents: a tool to explore complex dynamics}.
\newblock Cambridge University Press, 2016.

\bibitem[Poincar\'e(1899)]{Poincare1899}
H.~Poincar\'e.
\newblock \emph{Les méthodes nouvelles de la mécanique céleste. Tome III}.
\newblock GAUTHIER-VILLARS, 1899.

\bibitem[Potthast et~al.(2019)Potthast, Walter, and
  Rhodin]{potthast2019localized}
R.~Potthast, A.~Walter, and A.~Rhodin.
\newblock A localized adaptive particle filter within an operational {NWP}
  framework.
\newblock \emph{Monthly Weather Review}, 147\penalty0 (1):\penalty0 345--362,
  2019.

\bibitem[Raanes et~al.(2015)Raanes, Carrassi, and Bertino]{raanes2015extending}
P.~N. Raanes, A.~Carrassi, and L.~Bertino.
\newblock Extending the square root method to account for additive forecast
  noise in ensemble methods.
\newblock \emph{Monthly Weather Review}, 143\penalty0 (10):\penalty0
  3857--3873, 2015.

\bibitem[Raanes et~al.(2019)Raanes, Bocquet, and Carrassi]{raanes2019adaptive}
P.~N. Raanes, M.~Bocquet, and A.~Carrassi.
\newblock Adaptive covariance inflation in the ensemble kalman filter by
  gaussian scale mixtures.
\newblock \emph{Quarterly Journal of the Royal Meteorological Society},
  145\penalty0 (718):\penalty0 53--75, 2019.

\bibitem[Reinhold and Pierrehumbert(1982)]{reinhold1982dynamics}
B.~B. Reinhold and R.~T. Pierrehumbert.
\newblock Dynamics of weather regimes: Quasi-stationary waves and blocking.
\newblock \emph{Monthly Weather Review}, 110\penalty0 (9):\penalty0 1105--1145,
  1982.

\bibitem[Ruelle(1979)]{Ruelle1979}
D.~Ruelle.
\newblock Ergodic theory of differentiable dynamical systems.
\newblock \emph{Inst. Hautes \'Etudes Sci. Publ. Math.}, 50\penalty0
  (50):\penalty0 27--58, 1979.
\newblock ISSN 0073-8301.

\bibitem[Sakov and Bertino(2011)]{sakov2011relation}
P.~Sakov and L.~Bertino.
\newblock Relation between two common localisation methods for the enkf.
\newblock \emph{Computational Geosciences}, 15\penalty0 (2):\penalty0 225--237,
  2011.

\bibitem[Sakov and Oke(2008{\natexlab{a}})]{sakov2008deterministic}
P.~Sakov and P.~R. Oke.
\newblock A deterministic formulation of the ensemble {K}alman filter: an
  alternative to ensemble square root filters.
\newblock \emph{Tellus A}, 60\penalty0 (2):\penalty0 361--371,
  2008{\natexlab{a}}.

\bibitem[Sakov and Oke(2008{\natexlab{b}})]{sakov2008implications}
P.~Sakov and P.~R. Oke.
\newblock Implications of the form of the ensemble transformation in the
  ensemble square root filters.
\newblock \emph{Monthly Weather Review}, 136\penalty0 (3):\penalty0 1042--1053,
  2008{\natexlab{b}}.

\bibitem[Sakov et~al.(2018)Sakov, Haussaire, and Bocquet]{sakov2018iterative}
P.~Sakov, J.~M. Haussaire, and M.~Bocquet.
\newblock An iterative ensemble {K}alman filter in presence of additive model
  error.
\newblock \emph{Quarterly Journal of the Royal Meteorological Society}, 2018.

\bibitem[Shimada and Nagashima(1979)]{shimada1979}
I.~Shimada and T.~Nagashima.
\newblock A numerical approach to ergodic problem of dissipative dynamical
  systems.
\newblock \emph{Progress of Theoretical Physics}, 61\penalty0 (6):\penalty0
  1605--1616, 1979.

\bibitem[Snyder(1996)]{snyder1996summary}
C.~Snyder.
\newblock Summary of an informal workshop on adaptive observations and fastex.
\newblock \emph{Bulletin of the American Meteorological Society}, 77\penalty0
  (5):\penalty0 953--961, 1996.

\bibitem[Snyder et~al.(2008)Snyder, Bengtsson, Bickel, and
  Anderson]{snyder2008obstacles}
C.~Snyder, T.~Bengtsson, P.~Bickel, and J.~Anderson.
\newblock Obstacles to high-dimensional particle filtering.
\newblock \emph{Monthly Weather Review}, 136\penalty0 (12):\penalty0
  4629--4640, 2008.

\bibitem[Snyder et~al.(2015)Snyder, Bengtsson, and
  Morzfeld]{snyder2015performance}
C.~Snyder, T.~Bengtsson, and M.~Morzfeld.
\newblock Performance bounds for particle filters using the optimal proposal.
\newblock \emph{Monthly Weather Review}, 143\penalty0 (11):\penalty0
  4750--4761, 2015.

\bibitem[Szunyogh et~al.(2002)Szunyogh, Toth, Zimin, Majumdar, and
  Persson]{szunyogh2002propagation}
I.~Szunyogh, Z.~Toth, A.~V. Zimin, S.~J. Majumdar, and A.~Persson.
\newblock Propagation of the effect of targeted observations: The 2000 winter
  storm reconnaissance program.
\newblock \emph{Monthly weather review}, 130\penalty0 (5):\penalty0 1144--1165,
  2002.

\bibitem[Thompson(1957)]{Thompson1957}
P.~D. Thompson.
\newblock Uncertainty of initial state as a factor in the predictability of
  large scale atmospheric flow patterns.
\newblock \emph{Tellus}, 9\penalty0 (3):\penalty0 275--295, 1957.
\newblock \doi{10.1111/j.2153-3490.1957.tb01885.x}.

\bibitem[Tondeur et~al.(2020)Tondeur, Carrassi, Vannitsem, and
  Bocquet]{tondeur2020temporal}
M.~Tondeur, A.~Carrassi, S.~Vannitsem, and M.~Bocquet.
\newblock On temporal scale separation in coupled data assimilation with the
  ensemble kalman filter.
\newblock \emph{Journal of Statistical Physics}, 179:\penalty0 1161--1185,
  2020.

\bibitem[Toth and Kalnay(1993)]{toth1993ensemble}
Z.~Toth and E.~Kalnay.
\newblock Ensemble forecasting at nmc: The generation of perturbations.
\newblock \emph{Bulletin of the american meteorological society}, 74\penalty0
  (12):\penalty0 2317--2330, 1993.

\bibitem[Toth and Kalnay(1997)]{toth1997ensemble}
Z.~Toth and E.~Kalnay.
\newblock Ensemble forecasting at {NCEP} and the breeding method.
\newblock \emph{Monthly Weather Review}, 125\penalty0 (12):\penalty0
  3297--3319, 1997.

\bibitem[Trevisan and Legnani(1995)]{trevisan1995transient}
A.~Trevisan and R.~Legnani.
\newblock Transient error growth and local predictability: A study in the
  lorenz system.
\newblock \emph{Tellus A}, 47\penalty0 (1):\penalty0 103--117, 1995.

\bibitem[Trevisan and Palatella(2011)]{trevisan2011kalman}
A.~Trevisan and L.~Palatella.
\newblock On the {K}alman filter error covariance collapse into the unstable
  subspace.
\newblock \emph{Nonlinear Processes in Geophysics}, 18\penalty0 (2):\penalty0
  243--250, 2011.

\bibitem[Trevisan and Uboldi(2004{\natexlab{a}})]{trevisan2004}
A.~Trevisan and F.~Uboldi.
\newblock Assimilation of standard and targeted observations within the
  unstable subspace of the observation-analysis-forecast cycle.
\newblock \emph{Journal of the Atmospheric Sciences}, 61:\penalty0 103--113,
  2004{\natexlab{a}}.

\bibitem[Trevisan and Uboldi(2004{\natexlab{b}})]{trevisan2004assimilation}
A.~Trevisan and F.~Uboldi.
\newblock Assimilation of standard and targeted observations within the
  unstable subspace of the observation--analysis--forecast cycle system.
\newblock \emph{Journal of the Atmospheric Sciences}, 61\penalty0 (1):\penalty0
  103--113, 2004{\natexlab{b}}.

\bibitem[Trevisan et~al.(2010)Trevisan, D'Isidoro, and Talagrand]{trevisan2010}
A.~Trevisan, M.~D'Isidoro, and O.~Talagrand.
\newblock Four-dimensional variational assimilation in the unstable subspace
  and the optimal subspace dimension.
\newblock \emph{Q. J. R. Meteorol. Soc.}, 136:\penalty0 487--496, 2010.

\bibitem[Uboldi and Trevisan(2006)]{uboldi2006}
F.~Uboldi and A.~Trevisan.
\newblock Detecting unstable structures and controlling error growth by
  assimilation of standard and adaptive observations in a primitive equation
  ocean model.
\newblock \emph{Nonlinear Processes in Geophysics}, 16:\penalty0 67--81, 2006.

\bibitem[Uboldi et~al.(2005)Uboldi, Trevisan, and Carrassi]{Uboldi-2005}
F.~Uboldi, A.~Trevisan, and A.~Carrassi.
\newblock Developing a dynamically based assimilation method for targeted and
  standard observations.
\newblock \emph{Nonlinear Processes in Geophysics}, 12\penalty0 (1):\penalty0
  149--156, 2005.
\newblock \doi{10.5194/npg-12-149-2005}.

\bibitem[Vallis(2017)]{vallis_2017}
G.~K. Vallis.
\newblock \emph{Atmospheric and Oceanic Fluid Dynamics: Fundamentals and
  Large-Scale Circulation}.
\newblock Cambridge University Press, 2 edition, 2017.
\newblock \doi{10.1017/9781107588417}.

\bibitem[Vannitsem(2017)]{vannitsem2017predictability}
S.~Vannitsem.
\newblock Predictability of large-scale atmospheric motions: Lyapunov exponents
  and error dynamics.
\newblock \emph{Chaos: An Interdisciplinary Journal of Nonlinear Science},
  27\penalty0 (3):\penalty0 032101, 2017.

\bibitem[Vannitsem and Duan(2020)]{VannitsemDuan2020}
S.~Vannitsem and W.~Duan.
\newblock On the use of near-neutral backward lyapunov vectors to get reliable
  ensemble forecasts in coupled ocean–atmosphere systems.
\newblock \emph{Climate Dynamics}, 55:\penalty0 1125--1139, 2020.
\newblock \doi{10.1007/s00382-020-05313-3}.

\bibitem[Vannitsem and Lucarini(2016)]{vannitsem2016}
S.~Vannitsem and V.~Lucarini.
\newblock Statistical and dynamical properties of covariant lyapunov vectors in
  a coupled atmosphere-ocean model—multiscale effects, geometric degeneracy,
  and error dynamics.
\newblock \emph{Journal of Physics A: Mathematical and Theoretical},
  49\penalty0 (22):\penalty0 224001, 2016.

\bibitem[Vannitsem et~al.(2015)Vannitsem, Demaeyer, {De Cruz}, and
  Ghil]{vannitsemetal2015}
S.~Vannitsem, J.~Demaeyer, L.~{De Cruz}, and M.~Ghil.
\newblock Low-frequency variability and heat transport in a low-order nonlinear
  coupled ocean–atmosphere model.
\newblock \emph{Physica D: Nonlinear Phenomena}, 309:\penalty0 71 -- 85, 2015.
\newblock ISSN 0167-2789.

\end{thebibliography}

\end{document}